\documentclass[preprint2]{aastex}

\usepackage{graphicx}
\usepackage{amsmath} 
\usepackage{subfigure}
\usepackage{pdfpages}
\usepackage{longtable}
\usepackage{epsfig}
\usepackage{natbib}
\usepackage{color}
\usepackage{afterpage}
\usepackage{lscape}

\newcommand{\kms}{\mbox{${\rm km~s^{-1}}$}}

\newcommand{\Msun}{$M_{\odot}$}
\newcommand{\Lsun}{$L_{\odot}$}
\newcommand{\kmsMpc}{km s$^{-1}$ Mpc$^{-1}$}
\def\simgt{\lower.5ex\hbox{$\; \buildrel > \over \sim \;$}}
\def\simlt{\lower.5ex\hbox{$\; \buildrel < \over \sim \;$}}

\shorttitle{Galaxy Groups within 3500 km s$^{-1}$}
\shortauthors{Kourkchi \& Tully}

\begin{document}


\title{Galaxy Groups within 3500 km s$^{-1}$}


\author{Ehsan Kourkchi}
\affil{Institute for Astronomy, University of Hawaii, 2680 Woodlawn Drive, Honolulu, HI 96822, USA}
\and
\author{R. Brent Tully}
\affil{Institute for Astronomy, University of Hawaii, 2680 Woodlawn Drive, Honolulu, HI 96822, USA}

%



\begin{abstract}

A study of the group properties of galaxies in our immediate neighborhood provides a singular opportunity to observationally constrain the halo mass function, a fundamental characterization of galaxy formation.  Detailed studies of individual groups have provided the coefficients of scaling relations between a proxy for the virial radius, velocity dispersion, and mass that usefully allows groups to be defined over the range $10^{10} - 10^{15}$~\Msun.  At a second hierarchical level, associations are defined as regions around collapsed halos extending to the zero velocity surface at the decoupling from cosmic expansion.  The most remarkable result of the study emerges from the construction of the halo mass function from the sample.  At $\sim10^{12}$~\Msun\ there is a jog from the expectation Sheth-Tormen function, such that halo counts drop by a factor $\sim 3$ in all lower mass bins. 

\end{abstract}





\section{Introduction}
\label{chap:introduction}

Why construct another group catalog?  Consideration of our immediate vicinity is instructive.  It is said that we live in the Local Group.  In an N-body simulation, a group is synonymous with a virialized halo \citep{2009ApJS..182..608K}.  A halo on the largest scale of local collapse can be called a group because this `parent' halo can contain galaxies as `child' sub-condensations.  Accordingly, it is evident that the Local Group is a misnomer.  The Local Group contains two main halos, separately around the Milky Way and M31, each with an entourage of companions.  In addition there are several minor galaxies like NGC~6822, IC~1613, and WLM  that have not yet merged with either of the two main halos, so still are to be considered as the distinct residents of separate halos; ie, each one is its own little group.

All the nearby entities within roughly 1 Mpc are falling together.  In another Hubble time it will be legitimate to call this ensemble the Local Group.  We propose to call a region of infall an ``association".  In doing so, we are giving a tighter definition to an idea used in earlier discussions of groups \citep{1987ApJ...321..280T, 2006AJ....132..729T}.  Accordingly, we live in the Milky Way Group in the Local Association.  

The genesis of the present discussion lies in two recent papers about galaxy groups  \citep{2015AJ....149...54T, 2015AJ....149..171T}, hereafter T15$a$ and T15$b$.  The first of these papers gives attention to about a dozen groups and environments that have been studied in detail down to the inclusion of dwarfs as faint as $M_R \sim -11$.  The groups range in mass from the Coma Cluster at $\sim 10^{15}$~\Msun\ down to dwarf correlations at inferred group masses of $10^{11}$~\Msun.  It was found that groups (parent halos) across this wide range obey scaling relations between mass, velocity dispersion, and dimensions as anticipated from clustering theory.  The coefficients of the scaling relations are defined in this first paper.

In T15$b$ the scaling relations are applied to build a group catalog based on the 2MASS Redshift Survey quasi-complete to $K_s=11.75$ \citep{2012ApJS..199...26H}, hereafter 2MRS11.75, a compilation of 43,526 galaxies that provides panoramic coverage except for 9\% of the sky in the zone of obscuration.  The 2MASS catalog is exceptional because, within a flux limit which approximates a distance limit, it provides an essentially complete inventory of the important reservoirs of light.  The effective distance limit is roughly 130~Mpc ($H_0=75$~\kms~Mpc$^{-1}$) because beyond about that distance one looses $L^{\star}$ galaxies, so is only picking up the exponential cutoff tail of the Schechter function distribution \citep{1976ApJ...203..297S}.  Consequently, the upper limit for the domain of reliability of the 2MRS11.75 group catalog is taken to be 10,000~\kms.

There was a warning in T15$b$ that this 2MASS-based catalog is less than optimal within 3,000~\kms.  It is a minor concern that peculiar velocities can affect inferred intrinsic luminosities and spatial separations, hence the construction of groups.  The larger concern derives from the loss of low surface brightness flux from 2MRS11.75 (2MASS exposures were short and against a high infrared foreground).  Again there is a greater and lesser concern.  The lesser one is the loss of flux from the periphery of cataloged entries; ie, an undercounting of flux.  This problem can be statistically addressed.  However there is the greater concern that if a galaxy lacks a high surface brightness core then it is simply lost from 2MRS11.75.

A goal of the present paper is to explore the nature of the smallest groups.  T15$b$ ended with the presentation of the halo mass function for the 2MRS11.75 sample.  That mass function is considered unreliable below $8\times10^{12}$~\Msun.  It is evident that to do better we need to look nearby.  For example, we know that the Milky Way Group has many satellites, but at a distance beyond 3,000~\kms\ it would be seen as a single system (if properly parsed from M31).  From the many satellites we have good estimates of the virial dimension, velocity dispersion, and mass of the Milky Way Group.  If viewed from beyond 3,000~\kms\ those properties could only be inferred from the luminosity of the single system.  Small groups lacking a 2MRS11.75 presence would be entirely missed.  It will be seen further in the discussion that 2MRS11.75 provides a very sparse representation of local structure.

There are several ways to inventory the domain of low surface brightness galaxies.  Our interest will be in surveys that provide the widest possible sky coverage.  Observations in the 21~cm line of neutral Hydrogen are particularly useful because a large fraction of targets at low redshift are low surface brightness and the identification of the galaxy as nearby is confirmed by a velocity measurement that is usually good to 10~\kms.  Although most satellite dwarfs are typed dwarf spheroidal and found to be gas deficient, most dwarfs that are {\it not} satellites, the ones that will particularly interest us as markers of distinct halos, are typed dwarf irregular and typically are abundant in HI. The Large Magellanic Cloud with $M_{HI} = 4 \times 10^8$~\Msun\ would be detected at the extremity of our sample at 40~Mpc at a flux of $\sim 1$~Jy \kms. HI surveys that are particularly useful because of their wide coverage and access to low surface brightness systems include the early study that emphasized faint dwarfs by \citet{1981ApJS...47..139F} and the optically blind Parkes and Arecibo programs \citep{2001MNRAS.322..486B, 2011AJ....142..170H}.

A crucial contribution to our discussion comes from optical searches across the unobscured sky for faint dwarfs that might be nearby by \citet{1998A&AS..127..409K, 2000A&AS..146..359K}.  \citet{2013AJ....145..101K} provide a good compendium of nearby galaxies.  Subsequent imaging with Hubble Space Telescope has provided distance estimates for about 400 nearby galaxies from the apparent magnitude of the tip of  the red giant branch in each system \citep{2002A&A...383..125K, 2006AJ....131.1361K, 2009AJ....138..332J}.  With a group at, say, 4~Mpc, and a distance measurement accuracy of 5\% (zero point uncertainty aside), candidate members of a group can be located to within 200~kpc.  This dimension is comparable with the virial radius for a $10^{12}$~\Msun\ halo, leading to unambiguous group assignments.

Inevitably, coverage is incomplete spatially and seriously incomplete with distance.  Problems are greatest at halos below $10^{12}$~\Msun.  However there is a potential reward.  This faint end domain of the halo mass function has not to our awareness been explored observationally and can only be explored in our immediate neighborhood.

We return to comment on what was described above as a minor nuisance.  The use of redshifts to infer luminosities and derive group properties can sometimes lead to significant errors for nearby galaxies.  Two notably problematic regions are in the Leo Spur \citep{2015ApJ...805..144K}, where galaxies can be twice as far away as naively anticipated from their velocities because of a coherent flow pattern, and in close proximity to the Virgo Cluster, where galaxies have been captured by the cluster and velocities are decoupled from cosmic expansion.  We have measured distances for many but not all galaxies to be considered \citep{2013AJ....146...86T, 2016AJ....152...50T}.  A strategy will be discussed to deal with the ensuing mix of information. 

\section{Defining Groups}

In the framework of the standard cosmology, dark matter gravitationally collapses into halos that potentially hold baryonic matter in the form of galaxies. Halos can be merged together to build up more massive halos. In this picture, galaxy groups and clusters are the final products of the hierarchical mass assembly of galaxies embedded in massive dark matter halos. There has been no consensus among observers about how to define a group. There are various algorithms to identify groups of galaxies in a sample of galaxies limited by distance or apparent magnitude, given measured distances and/or radial velocities.

Algorithms developed to find galaxy groups in various radial velocity and mass ranges fall in two main categories, friend-of-friend (FOF) and hierarchical. In FOF algorithms, galaxies are paired based on their proximity in radial velocity space and projected spatial separations. There have been many attempts using different variations of the FOF algorithm to identify groups 
\citep{1982ApJ...257..423H, 1983ApJS...52...61G, 1989ApJS...69..809M, 2000ApJ...543..178G, 2002AJ....123.2976R, 2002MNRAS.335..216M, 2004MNRAS.348..866E, 2007ApJ...655..790C, 2011MNRAS.416.2840L}.
These studies generally do not consider physical scaling characteristics of groups. For example, the mass of a group defines separation 
scales and velocity differentials.  In a FOF based recipe,  \citet{2011MNRAS.412.2498M} do introduce a mass effect with an assumption about how $K_s$-band luminosity and mass are correlated. They present a catalog of nearby galaxy groups for the volume within 3,000 km s$^{-1}$. 

Hierarchical algorithms associate galaxies in a tree construction process usually drawing on luminosities \citep{1978A&A....63..401M, 1984TarOT..73....1V, 1987ApJ...321..280T, 1987MNRAS.225..505N, 1993ApJS...85....1N}.  An interesting feature of hierarchical methods is the recovery of a quasi-continuum of linkage levels.  One can decide on the level that defines a group after the construction of the hierarchy.  A catalog might record multiple levels; e.g., separate levels identified with ``groups" and ``associations" \citep{1987ApJ...321..280T}.

A distinctive approach is to define groups in accordance with anticipated scaling relations.  One variant is to base expectations on N-body simulations 
\citep{2005MNRAS.356.1293Y, 2006MNRAS.366....2W}.  The variation presented in T15$b$ used scaling relations grounded in direct observations.  The present study finds application of the T15$b$ methodology with a local sample of galaxies.  Multiple clustering levels can be identified with this procedure because scaling relations are separately available for the conditions of virialization and the onset of collapse.


Galaxy linkages are a challenge with limited knowledge about their spatial distributions.  Projected positions are well known but distance information provides useful discrimination in only exceptional circumstances (very nearby or, to a degree, in the Virgo Cluster).  Velocities give only crude information.  The Hubble relation breaks down in close proximity to groups and interlopers can lurk within the range of group velocity dispersions.  The problem gets worse as distance increases and the number of identified members of groups diminishes so only more massive/luminous groups are found. 

All algorithms are vulnerable to errors in group assignments.  The following summarizes our ambition in generating a new group catalog of the local volume.

\begin{itemize}
\item The groups should manifest properties that sensibly scale in agreement with well observed examples over the mass range $10^{11}$ to $10^{15}$~\Msun.
\item Locally, aside from near the Virgo Cluster, velocity dispersions are modest so a correlation in velocities is a good indicator of the existence of a group.
\item Distances exist for many galaxies, some quite constraining, and this information gives clarification to absolute luminosities and scales.  However, individual distances may be misleading because of errors.  A mixing of measured distances and redshift distances can lead to large scatter.  A nuanced use of distance information is required.
\item  Most large groups will contain galaxies contained within the 2MRS11.75 catalog and the masses of these groups can be expected to align with those identified in T15$b$.  However small groups found nearby have poorly determined infrared fluxes and in these cases the linkage between observed light and mass has considerable uncertainty.
\item  The immediate region around the Virgo Cluster is particularly confusing because of projection effects and velocity streaming.  Distance measurements provide some insight.  The identification of separate groups involves judgement and probably errors are made.
\end{itemize}

The outline of this paper is as follows. In \S \ref{sec:sample}, we will describe our galaxy sample. In \S \ref{sec:general_group}, we will review the general properties of several well-studied galaxy groups that will be used as inputs to our group finding algorithm explained in \S \ref{sec:algorithm}. Special cases are discussed in \S \ref{sec:virgo} - \S \ref{sec:10Mpc}.  Our catalog of the identified galaxy groups will be presented in \S \ref{sec:group.catalog}. In \S \ref{subsec:investigations} we examine the properties of the identified groups. In \S \ref{sec:massFunction} we present the group mass function in the local universe and finally in \S \ref{summary} there is a summary of our results. Throughout this paper, we assume the value of $H_0=75$ km s$^{-1}$ Mpc$^{-1}$ for the Hubble constant, consistent with the scale of direct distance measurements \citep{2016AJ....152...50T}.  

\section{Galaxy sample}
\label{sec:sample}

Our group identification algorithm requires a good estimation of the mass of galaxies. Stars are the dominant baryonic component of 80\% galaxies \citep{2016AN....337..306K}.  Stellar luminosity can be used as a proxy for the total mass of a galaxy \citep{2001ApJ...550..212B}.  The near-infrared flux is less affected by dust and the stochastic nature of young blue stars of a galaxy so it is reasonable to use the $K_s-$band luminosity of a galaxy as its stellar mass indicator.\footnote{The $K_s$ filter blocks the worst of foreground thermal emission in the $K$-band atmospheric window.}

We give consideration to galaxies with $V_{LS}<4000$ km s$^{-1}$, where $V_{LS}$ is the radial velocity relative to the Local Sheet \citep{2008ApJ...676..184T}, a variant on the more familiar but ambiguous Local Group reference frame  \citep{1977ApJ...217..903Y, 1996AJ....111..794K, 1999AJ....118..337C}. Our intent is to capture groups (down to groups of 1) within $V_{LS}=3500$~\kms.  The initial inclusion of galaxies with larger velocities is designed to achieve completion within groups.  The 2MASS Redshift Survey, 2MRS11.75 \citep{2012ApJS..199...26H} provides radial velocities for 98\% of galaxies in the 2 Micron All Sky Survey \citep{2000AJ....119.2498J, 2003AJ....125..525J} at Galactic latitudes greater than $5^{\circ}$ and brighter than $K_s=11.75$ mag. At the outer limit of our survey, a galaxy at the faint limit of the 2MRS11.75 survey has absolute luminosity $M_{K_s}=-21.6$, 2 magnitudes fainter than $M_{K_s}^{\star}=-23.55$ (T15$b$), the characteristic inflection point of the Schechter function \citep{1976ApJ...203..297S}.  Outside the Galactic gap of $\pm 5^{\circ}$, the 2MRS11.75 survey includes a substantial majority of baryonic luminosity within the study volume.  We supplement with catalogs identified below that contribute low surface brightness systems.
 Although the masses of these faint galaxies do not make an important contribution to the total mass of big halos, the inclusion of these galaxies improves statistics when deriving the dynamical parameters of groups.  More important, most of the lowest mass halos contain {\it only} low surface brightness systems.  A proper inventory of the halo mass function requires inclusion of small and faint galaxies.
 
Consequently we have given attention to:
\begin{itemize}

\item The Lyon-Meudon Extragalactic Database (LEDA\footnote{\url{http://leda.univ-lyon1.fr/}})

\item The \citet{2011MNRAS.412.2498M} galaxy group catalog with limiting velocity of V$_{LG} <$ 3,500 km s$^{-1}$ (Local Group reference frame). This catalog involves a compilation of galaxies taken from 2MASS, LEDA and NED\footnote{NED: The NASA/IPAC Extragalactic Database:\\ \url{https://ned.ipac.caltech.edu/}} databases.

\item The Updated Nearby Galaxy Catalog \citep{2013AJ....145..101K}, of 825 galaxies considered to be within 11 Mpc.

\item An augmentation of John Huchra's `Zcat', circa 2002\footnote{https://www.cfa.harvard.edu/$\sim$dfabricant/huchra/zcat/} that has been privately compiled and made available at  the Extragalactic Distance Database.\footnote{http://edd.ifa.hawaii.edu}

\item Distance measurements taken from the Cosmicflows-3 (CF3) catalog \citep{2016AJ....152...50T}.
\end{itemize}

We have compiled a sample of 17,491 galaxies with $V_{LS}<4000$ km s$^{-1}$,  that includes 3,300 with Cosmicflows-3 distance information.  The introduction of candidates from heterogeneous sources introduces a risk of inclusion of spurious objects (stars, galaxy shredding, erroneous redshifts, duplications due to small position differences).  Potential sources within 1000~\kms\ have been visually evaluated.  It might be argued that only candidates within well defined limits should be considered (say, flux limits).    However experience has shown that specialized surveys identify unusual objects.  HI surveys can recover low Galactic latitude and almost invisible low surface brightness galaxies \citep{1981ApJS...47..139F, 2016AJ....151...52S}; ultraviolet or spectroscopic observations can locate star forming compact dwarfs \citep{1989SoSAO..62....5M}; redshift surveys that include supposedly stellar objects identify ultra compact dwarfs \citep{2000PASA...17..227D, 2011ApJ...737...86C}; wide field imaging leads to the identification of resolved stellar systems that otherwise are not apparent \citep{2005ApJ...626L..85W, 2013ApJ...772...15M}.  However inevitably the investigations that lead to such serendipitous discoveries cover only parts of the sky, sometimes very small parts.  We do not want to ignore this information.  However making use of heterogeneous data is a challenge.


We infer stellar masses from $K_s$ magnitudes.
In case of a lack of $K_s$-band photometry, we use the $B-K_s$ color index together with the galaxy morphological type to convert $B$-band magnitudes to $K_s$-band magnitudes based on the following relations discussed by \citet{2003AJ....125..525J} and \citet{2005AstL...31..299K}.

\begin{eqnarray}
\langle B-K_s\rangle &=& +4.10 \qquad \textrm{$T\leq2$, (E, S0, Sa),} \nonumber\\
\langle B-K_s\rangle &=& +2.35 \qquad \textrm{$T\geq9$ (Sm, Im, Irr),}  \nonumber \\
\langle B-K_s\rangle &=& 4.60-0.25 T \qquad \textrm{$T=3\textrm{--}8$},\nonumber
\end{eqnarray}
where, $T$ is the digital galaxy morphological type in the RC3 catalog \citep{1991trcb.book.....D}. We take the $B$-band magnitudes from LEDA or NED catalogs. Other complementary information regarding our sample galaxies is taken from the Extragalactic Distance Database (EDD\footnote{\url{http://edd.ifa.hawaii.edu/dfirst.php}}).


\section{General properties of groups}
\label{sec:general_group}

Groups can be defined as structures with enough accumulated mass to have decoupled from the Hubble flow and collapsed.  However, it is not always easy for observers to define a group. Two natural radii that define the size of gravitationally bound systems can be applied with N-body simulations. One is $r_{200}$, the radius that encloses a region with an overdensity $200$ times larger than the ``critical density" for a matter dominated closed universe, $\rho_{crit}=3H^2_0/8\pi G$, and the other is the dynamical virial radius of a system in gravitational equilibrium, $r_g$,

\begin{equation}
r_g = \frac{\sum\nolimits_{i>j} ~ m_i m_j}{\sum\nolimits_{i>j} ~ m_i m_j /r_{ij}},
\label{eq:virial}
\end{equation}
where $m_i$ is the mass of $i^{th}$ galaxy and $r_{ij}$ is the separation distance between $i^{th}$ and $j^{th}$ galaxies. 

However neither $r_{200}$ nor $r_g$ is very useful to observers if the number of group candidates is small, possibly even just one galaxy. Other more observationally useful dimensions can be considered. With the collapse of a spherically symmetric over-dense region embedded in the expanding universe, the collapse time, $t_c$, is related to density by $t_c \propto \rho^{-1/2}$. Equating $t_c$ to the age of the Universe, one identifies a common density for all structures at the same phase of collapse (within the spherical approximation).   For example, there is a density and an associated mass-dependent radius of decoupling from cosmic expansion that can be called the first turnaround radius, $r_{1t}$, or the ``surface of zero-velocity" bounding a region of collapse.

The collapse of regions of higher density began at earlier times. There is a density associated with regions that collapsed, re-expanded, and by today have just reached turnaround in anticipation of re-collapse.   This mass-dependent radius can be called the second turnaround radius, $r_{2t}$. In clean cases as discussed in T15$a$, the caustic of second turnaround can be identified from a density drop and/or a transition in velocities from infall to virialization.

T15$a$ presented a detailed study of 8 groups with sizes and masses ranging from M31 to the Coma Cluster. Scaling relations based on that study inform our group identification algorithm.  According to T15$a$, there is a linear relationship between the two measured parameters

\begin{equation}
\sigma_p/R_{2t} = 368 ~ \rm{km~s^{-1}~Mpc^{-1}}
\label{eq:sig_R2t}
\end{equation}
where $R_{2t}$ is the projected second turnaround radius of a group with the line-of-sight velocity dispersion $\sigma_p$ given by

\begin{equation}
\sigma_p = \sqrt{\sum\limits_{i}(v_i-\langle v \rangle)^2/N)}~,
\label{eq:sigma-p}
\end{equation}
with $\langle v \rangle$ the group mean velocity and $N$ the number of members.  By convention, we use lowercase $r$ for a 3-dimensional radius and an upper case $R$ for a projected radius.  Here, dimensions are based on direct distance measurements with consistency between T15$a$ and the scale used in the present study.  It is to be noted that distances in the companion paper T15$b$ were based on velocities, such that the scaling relation coefficients varied with dependences on $h=H_0/100$.

A minor point to note is that
the value of $r_g$ in Eq.~\ref{eq:virial} is dominated by a small number of massive galaxies. Therefore, the virial radius is more practically defined in an unweighted form as 

\begin{equation}
R_g = \frac{N^2}{\sum\nolimits_{i>j} ~ 1 /R_{ij}}~,
\label{eq:virial_unweight}
\end{equation}
where $R_g$ is the projected group radius, $R_{ij}$ is the projected distance between galaxy pairs, $N$ is the number of group candidates and each pair is counted only once.  The group virial mass, $M_v$ is given by

\begin{equation}
M_v = \sigma^2_{3D} r_g / G = (\alpha \pi /2G) \sigma^2_p R_g~,
\label{eq:virial_mass}
\end{equation}
where $\sigma_{3D}$ is the 3-dimensional velocity dispersion, $\sigma_{3D}=\sqrt{\alpha} \sigma_p$.  The parameter $\alpha$ depends on the degree of virialization of orbits and following T15$a$ we assume $\alpha=2.5$.
The following statistical correlations from T15$a$ are assumed: $r_{2t}=\sqrt{3/2}R_{2t}$, $r_g=(\pi/2)R_g$ and $r_g = 1.05r_{2t}$.


T15$a$ also demonstrates the correlation

\begin{equation}
R_{2t} = 0.215~M^{1/3}_{12}~,
\label{eq:R_Mv}
\end{equation}
where $M_{12} = M_v/10^{12} M_{\odot}$.   Equations \ref{eq:sig_R2t} and \ref{eq:R_Mv} can be combined to give

\begin{equation}
M_v = 2.0 \times 10^6 \sigma^3_p~.
\label{eq:M_v.Sigma}
\end{equation}

Given an estimate of the virial mass of a group, $M_v$, Eq.~\ref{eq:R_Mv} gives an estimate of the projected second turnaround radius and Eq.~\ref{eq:M_v.Sigma} gives the expectation dispersion.  In a group searching algorithm, galaxies can be associated to groups of a given mass if they lie within the anticipated projected radius $R_{2t}$ and, say, twice the anticipated velocity dispersion $\sigma_p$. 

A final critical ingredient is an assumption about the mass-to-light ratio of halos, $M_v/L$.   Given a total absolute $K_s$-band luminosity we want to infer a group mass, whence a group dimension and velocity dispersion.  There is the complication that $M/L$ is not a constant value but rather increases with group mass over the familiar domain of groups and clusters \citep{2002ApJ...569..101M, 2005ApJ...618..214T, 2009ApJ...695..900Y}.  Mass per unit light is minimal in halos with $L_{K_s} \sim 10^{10}$~\Lsun.  Faintward, luminosity falls off faster than mass.  The details at the low luminosity end are poorly constrained.

\begin{figure}[!]
\begin{center}
\includegraphics[width=8cm]{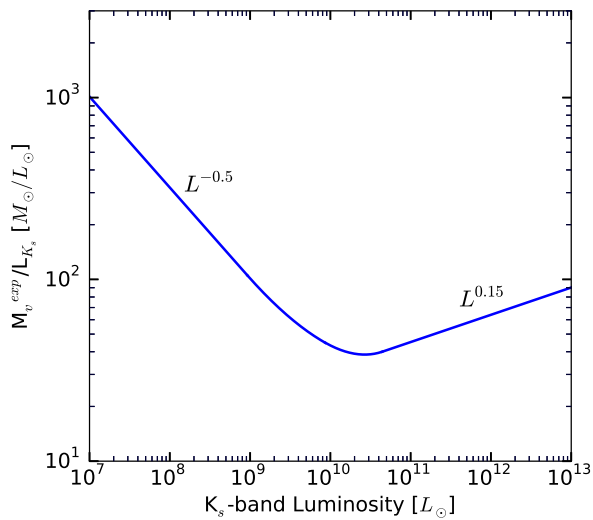}
\caption{
Assumed relation between the expected virial mass and $K_s$ luminosity of halos.}
\label{fig:massTolight}
\end{center}
\end{figure}

We adopt the following formulation for the expected infrared mass-to-light ratio, $M^{exp}_v / L_{K_s}$.  The dependence above $10^{10}$~\Lsun\ is justified in T15$b$, with an appropriate conversion of the distance scale with $h=0.75$.  The implications for the description at fainter luminosities are considered further in Section~\ref{sec:10Mpc}.

\begin{equation}
\label{eq:M_L}
\frac{M^{exp}_v}{L_{K_s}} = \left\{
\begin{array}{l l}
   32\times L_{10}^{-0.5} & L_{10} < 0.0927   \\
   32\times L_{10}^{0.15} &  L_{10} >  4.423 
   \\
\end{array} \right.
\end{equation}
where $L_{10}$ is the $K_s$ luminosity in units of $10^{10} L_{\odot}$ and $M^{exp}_v / L_{K_s}$ is in units of $M_{\odot}/L_{\odot}$. To have a smooth transition for $0.0927 \leq L_{10} \leq 4.423$, we connect the low and high luminosity relations using two circular curves as seen in Figure \ref{fig:massTolight}.

\section{Galaxy Associations}
\label{sec:assoc}

The projected radius of second turnaround $R_{2t}$ is an attractive parameter because it approximates the virial radius and has observable manifestations.  The projected radius of first turnaround $R_{1t}$ that prescribes the zero-velocity surface separating expansion from collapse is also of interest.  Exquisite distance resolution is required to observationally define zero-velocity surfaces.  T15$a$ discusses three reasonably established cases, those enclosing our local region, the M81 Group, and the Virgo Cluster respectively.  It was concluded in that reference that $r_{1t}/r_{2t} = 3.14 \pm 0.28$, a value in reasonable agreement with the prediction of the standard $\Lambda$CDM model $r_{1t}/r_{2t} = R_{1t}/R_{2t} \sim 3.5$ if $\Omega_{\Lambda} = 0.7$ 

The identification of aggregations of galaxies on larger scales than collapsed halos has some currency \citep{1975gaun.book..557D, 2007ApJ...655..790C}  and the term `association' has been used \citep{1987ApJ...321..280T,2006AJ....132..729T, 2011MNRAS.412.2498M}.  We propose that, while groups/clusters/nests be defined as regions of collapse with dimensions $r_{2t}$, associations be defined as regions of infall with dimensions $r_{1t}$.

Accordingly, to take the region where we live as an example, we live in the Milky Way Group and are linked with the M31 Group and the halos of a known $\sim$dozen dwarfs in the Local Association, a domain of radius $\sim 1$~Mpc.  The Virgo Cluster has an $R_{2t}$ projected radius of 1.9 Mpc while the Virgo Association, enclosing the infall region, has a radius of 7 Mpc.

Another radius of physical interest is the zero energy surface enclosing the domain of ultimate collapse.  At present this radius can only be inferred from a cosmological model.  It is expected to be $\sim 40\%$ larger than $R_{1t}$ \citep{2008A&A...488..845P}.  The Virgo Cluster is 16 Mpc from us.  The front edge of the current infall region lies about 9 Mpc from us.  Galaxies as close as $\sim 6$~Mpc from us can expect to ultimately fall into the cluster but we will not.  Probably, though, the Local, Maffei-IC342, and M81 associations will all fall together.  The CenA-M83 and NGC 253 associations can expect to remain isolated. 

\section{The group-finding algorithm}
\label{sec:algorithm}

The region projected onto the Virgo Cluster and its immediate surroundings is a special case and will be given attention in a separate section. The radius of second turnaround is taken to be at $6.8^{\circ}$ centered on M87 and an annulus of considerable confusion extends out another $\sim5^{\circ}$.

Setting aside the Virgo core, the algorithm we use here is basically the procedure described in T15$b$ with modifications compelled by the need to reconcile candidates with measured distances with errors and other candidates with known velocities but not distances. Initially, distances will be inferred from velocities ($H_0=75$~\kms~Mpc$^{-1}$) because distances are usually unknown and, when known, individual distance uncertainties can be large.  As a second step, distance information is evaluated, giving particular attention to groups with multiple measures that provide low uncertainties and to measures, particularly by the Cepheid, supernova, and tip of the red giant branch methods that provide low individual uncertainties.  If a distance based on a redshift deviates by more than an estimated one-sigma uncertainty in a measured distance then the measured distance is preferred.

Here is the decision sequence.

1) Intrinsic luminosities are assigned to each member of the full sample of 17 thousand galaxies based on apparent $K_s$ luminosities and velocities assuming $H_0 = 75$~\kmsMpc.  Masses are assigned using Eq.~\ref{eq:M_L}.

2) Starting with the intrinsically most luminous member of the sample, we search for all other galaxies that lie within $R_{2t}$ and $2\sigma_p$ of this galaxy.  The mass assignments from luminosities define $R_{2t}$ and $\sigma_p$ through Eqs.~\ref{eq:R_Mv} and \ref{eq:M_v.Sigma}.
We now have a tentative group with luminosity equal to the sum of the parts, at a luminosity weighted location and with an unweighted average velocity.

3) The search is repeated with the expanded parameters for the tentative group.  If new candidate members are found then revised luminosity, spatial, and velocity properties are determined and the cycle is repeated as often as necessary until no new linkages are made.

4) The procedure is repeated with the next most luminous galaxy not yet assigned to a group, and with sequentially less luminous systems until every galaxy in the sample has been considered.

5) Once a tentative list of groups has been compiled, attention is given to the possible overlap of some groups in position and velocity.  Two spatially overlapping groups are merged if either their radial velocity differences are less than three times their maximum velocity dispersion or both have consistent measured distances.  If groups are merged then once again the global group parameters are determined and the sample is scanned to check if new linkages have arisen.  Again, repeat cycles might be required until there are no changes.

6) Some confusing cases were manually investigated to remove interlopers and/or to add new possible group members. Examples are given among the special cases in \S8 and \S9.  Once distance information is incorporated, renewed cycles lead to final adjustments.

Once the group catalog is finalized, the distance of each group is estimated based on the weighted average of CF3 measured distances of its members. If none of the members has distance measurements, we use Hubble distances. 

To find the group associations, we follow a similar procedure to that described above to link galaxy groups and ungrouped individual galaxies into associations. The search and merge process is performed in 3 dimensional spatial space using the first turnaround radii, $r_{1t}$. All halos within a common association are absorbed.  Overlapping associations are merged into larger associations, with each step followed by an update of characterizing parameters. The process is repeated until no more associations can be merged.

\section{Special Case: The Virgo Cluster}
\label{sec:virgo}

\begin{figure}[!]
\begin{center}
\includegraphics[width=3.1in]{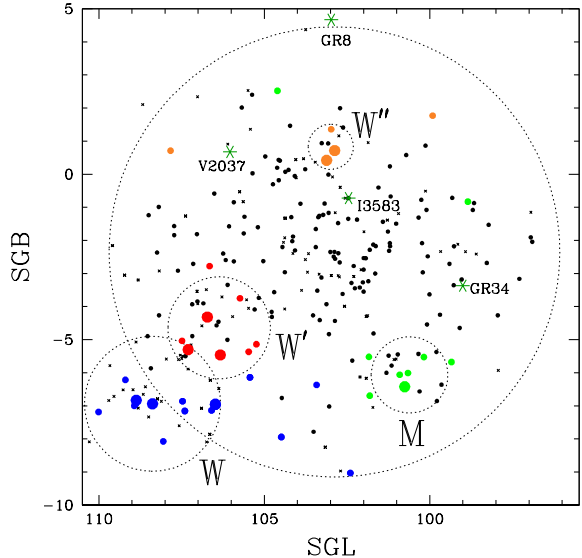}
\caption{
Projection confusion in the Virgo Cluster.  Small black symbols identify 2MRS11.75 galaxies within the $6.8^{\circ}$ radius of second turnaround of the Virgo Cluster or the $2^{\circ}$ radius domain of the Virgo~W Cluster.  Colored symbols identify galaxies with distance measurements that indicate that they are {\it not} in the Virgo Cluster; larger symbols for more accurately known distances.  Galaxies at roughly twice the Virgo distance around 34~Mpc are in blue if associated with Virgo~W and green if associated with the M cloud or more scattered.  Galaxies roughly 50\% farther than Virgo around 24 Mpc are in red if in Virgo~W$^{\prime}$ or in orange if in an entity labeled W$^{\prime\prime}$ or more scattered.  Galaxies identified by dark green asterisks are to the foreground; 3 of them at about 9 Mpc and the 4th at the edge of the cluster, GR8 at 2 Mpc.}
\label{fig:vir68}
\end{center}
\end{figure}

After accounting for Galactic rotation, the only galaxies in the sky with peculiar velocities sufficient to cause blueshifts either lie within 1~Mpc or are projected against the Virgo Cluster.   The Virgo dwarf VCC~846 has $V_{LS}=-821$~\kms.  Galaxies with velocities as high as 2800~\kms\ have been attributed to the cluster.  In constructing a catalog of nearby groups, velocities are not a useful indicator of distance over an area of 145 sq. deg. of sky centered on the giant elliptical M87.  It is a particular concern that a casual inspection reveals there is an important group/cluster projected at the edge of the Virgo Cluster; the entity slightly to the west called Virgo W  \citep{1961ApJS....6..213D}.  Virgo~W has a group velocity of $V_{LS}=2261$~\kms, lies at a distance of 34~Mpc (measures in CF3 for 18 galaxies), has velocity dispersion 340~\kms, and mass approximating $10^{14}$~\Msun.  Virgo~W is comparable in mass to the Fornax Cluster and its virial radius abuts the virial radius of Virgo in projection.  The proximity of the two entities in projection is seen in Figure~\ref{fig:vir68}.

There is convincing evidence that another structure called Virgo~W$^{\prime}$ is 50\% farther away than the Virgo Cluster although entirely confused in projection and velocities.   In a study of distances using the surface brightness fluctuation technique with Hubble Space Telescope, five galaxies were found to be distinctly more distant than the rest of the sample \citep{2007ApJ...655..144M, 2009ApJ...694..556B}.  These results are confirmed by Cepheid and supernova Type I$a$ observations of NGC~4639 \citep{2001ApJ...553...47F, 2007ApJ...659..122J}.  The Virgo~W$^{\prime}$ Group, with $V_{LS}=921$~\kms, is at 23.7~Mpc, 7.7~Mpc directly behind the Virgo Cluster. 

Several galaxies have been demonstrated to lie to the foreground of the cluster from tip of the red giant branch measurements of stars resolved in Hubble Space Telescope images \citep{2014ApJ...782....4K}.  GR34 (9.3~Mpc), IC~3583 (9.5~Mpc), and VCC~2037 (9.6~Mpc) project directly onto the cluster as does the previously known GR8 (2.1~Mpc).

In addition to these compelling cases of projection, there are a number of suspected cluster interlopers.  Members of the \citet{1984ApJ...282...19F} M Cloud have distance estimates at 35~Mpc, associating this feature with Virgo~W.   The good distance estimates mentioned above alert us to expect contamination at depths of 9, 24, and 34 Mpc.  Undoubtedly other galaxies remain to be identified as Virgo interlopers.  The general properties of the Virgo Cluster can be established but the detailed properties of the projected structures are uncertain since major components might be mis-assigned in distance. 

\subsection{The Virgo Association}

\begin{figure*}
\begin{center}
\includegraphics[width=17cm]{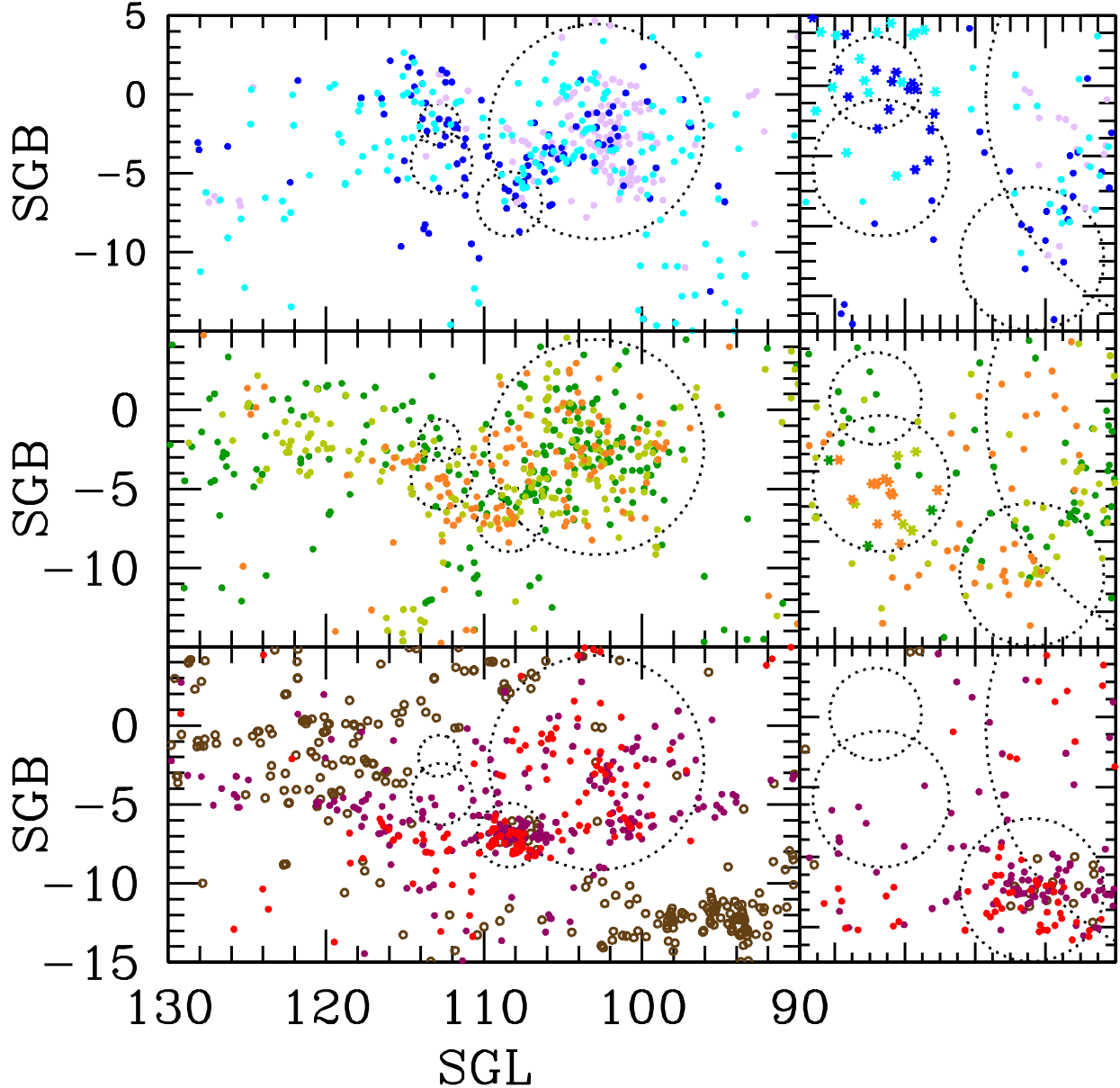}
\caption{
Clustering in the Virgo Southern Extension.  Galaxies plotted in the top panel have velocities $400<V_{LS}<1000$~\kms, in 200~\kms\ bins colored purple, blue, and cyan.  Galaxies in the middle panel lie in the range $1000-1750$~\kms\ in 250~\kms\ bins of dark and pea green and orange.  Galaxies in the bottom panel lie in the range $1750-3500$~\kms\ in bins of (red) 250, (mauve) 500, and (open brown) 1000~\kms.  Circles locate Virgo and Virgo W clusters as in Fig.~\ref{fig:vir68} and the PGC 41789 and PGC 42734 groups in the southern extension.  The immediate regions of the groups are enlarged in the right panels.}
\label{fig:vsx.map}
\end{center}
\end{figure*}

A larger region around the Virgo Cluster is seen in Figure~\ref{fig:vsx.map}.  See also the discussion by \citet{2016arXiv161100437K}.   For the moment, we draw attention to the concentration of galaxies to the left of the Virgo Cluster, a structure  \citet{1961ApJS....6..213D} has called the Virgo Southern Extension.  The region of infall around Virgo, the Virgo Association, extends to a radius $r_{1t}\sim7$~Mpc, or about $25^{\circ}$ in projection \citep{2014ApJ...782....4K}.    As the cluster is approached, departures from Hubble expansion become extreme.  The relationship between velocities and distances is complex because of projection effects.  Nevertheless discrete groups within the Southern Extension can be identified from spatial-velocity correlations, although detailed separations are uncertain because of overlaps.  

We draw attention to two particularly interesting features in Figure~\ref{fig:vsx.map}.  The three panels present the distribution of galaxies in and adjacent the Virgo Cluster in three velocity regimes with splits at 1000 and 1750~\kms.  The Virgo Cluster, within the largest circle, is well represented in all three velocity intervals.  Virgo~W shows up in the middle and bottom panels in the circle that infringes on the Virgo Cluster.  Virgo~W is embedded in a filament colored red and mauve extending over $94^{\circ}<SGL<120^{\circ}$ at $SGB \sim -7^{\circ}$.  Background features ($V_{LS}>2500$~\kms) are seen in the upper left and lower right parts of the bottom panel.

The interest with the present discussion is the regions of the two circles disjoint from the Virgo Cluster.  The upper one, at $SGL=112.68^{\circ}$, $SGB=-2.13^{\circ}$, is centered on a group with dominant galaxy PGC 42734 (NGC 4636).\footnote{Groups will be identified by the Principal Galaxies Catalog (PGC) identification of the most luminous member at $K_s$ band.  PGC names are numeric strings of up to 7 digits.}  We have 29 galaxies linked to the group, with a group velocity of $<V_{LS}> = 794$~\kms\ and a weighted average distance of $16.2\pm1.1$~Mpc (the uncertainty is derived from the sum of the weights).  The lower circle is centered at $112.78^{\circ}$, $-4.23^{\circ}$ with PGC 41789 the most prominent galaxy (NGC 4527).  In this case the group of 22 candidates has a velocity of 1550~\kms\ and the averaged distance is $14.3\pm0.7$~Mpc.  The PGC 42734 Group shows up prominently in the top panel while the PGC 41789 Group makes an appearance in the middle panel.

The Virgo Cluster itself (PGC 41220) has a velocity (averaged over 170 galaxies identified as members from distance measures) of $<V_{LS}>= 1058\pm56$~\kms\ and a weighted distance of $15.8\pm0.5$~Mpc.  Hence group PGC 41789 has a {\it higher} velocity by $+492$~\kms\ and has a distance that is {\it smaller} than that of the cluster while group PGC 42734 has a {\it lower} velocity by $-264$~\kms\ and nominally its distance is {\it larger} than that of the cluster.  This pattern is expected within the infall region.  The PGC 42734 entourage is falling toward us from the back side of the cluster while PGC 41789 and friends are falling away from us from the front side of the cluster.  The observed velocity field is exploited in a modeling of orbits giving an estimation of the cluster mass by Shaya et al. (in preparation).

\begin{figure}[!]
\begin{center}
\includegraphics[width=8cm]{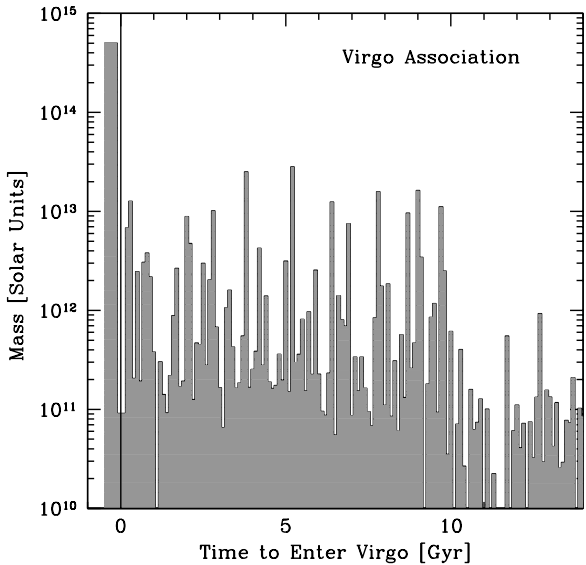}
\caption{Influx into the Virgo Cluster.  The histogram bins reflect the mass in galaxies that will arrive at the cluster second turnaround surface (roughly the virial radius) with time in the future.  The bar to the left of zero is set by the current mass within the cluster.
}
\label{fig:virgoinfall}
\end{center}
\end{figure}

The PGC 41789 and PGC 42734 groups are extreme examples of infall but, of course, these groups are being pursued by many more galaxies as they fall into the Virgo Cluster.  The concept of associations as the regions around halos that have decoupled from Hubble expansion was introduced in Section \ref{sec:assoc}.  By far the largest association within the volume under discussion has the Virgo Cluster at its core. \citet{1984ApJ...281...31T} provided a quantitative estimate of the infusion of galaxies into the cluster.   An update of that discussion is illustrated by Figure~\ref{fig:virgoinfall}.  Once a galaxy is within the zero velocity surface, the decoupling from cosmic expansion,  the time to full collapse is given in the spherical approximation by
\begin{equation}
t=(r_{1t}^3 /8GM(r))^{1/2}(\eta-{\rm sin} \eta)
\end{equation} 
where
\begin{equation}
{\rm cos} \eta = 1-2r/r_{1t}
\end{equation}
$r$ and $r_{1t}$ are the 3D radius of the galaxy from the cluster center and the first turnaround radius respectively, and M(r) is the mass interior to $r$.

With our current inventory, using measured distances as available and otherwise complemented by distances from velocities, we identify 1683 galaxies within the Virgo Association, 714 (40\%) of these already within the core defined by the second turnaround radius.  The galaxies in the association have an integrated $K_s$ band luminosity of $1 \times 10^{13}$~\Lsun\ with $5 \times 10^{12}$~\Lsun\ in the collapsed core (50\%).  By mass, there is $7 \times 10^{14}$~\Msun\ within the association, $4 \times 10^{14}$~\Msun\ already in the core (roughly 60\%).  Details are uncertain.  The galaxy counts are particularly crude because sampling is uneven in both distance and projection.  The luminosity figures are best known because large galaxies dominate and are easily seen across the region of interest.  The mass figures rely on assumptions susceptible to systematics.  As a point of comparison, the orbit modeling that will be reported by Shaya et al. prefer a mass for the Virgo core of $5 \times 10^{14}$~\Msun. 

\section{Special Cases: Velocity Anomalies}
\label{sec:velocityAnomalies}

The direction toward the Virgo Cluster is particularly challenging because of the complexity of the infall zone and the unfortunate near alignment of the background Virgo W Cluster and related filament.  There are several other difficult lines-of-sight that should be mentioned.  They tend to be of two types: either cases of groups with unexpectedly large internal velocity excursions or cases of projections where velocities are similar but distances are manifestly different.

A case of the first type occurs within the Virgo infall region.  Several galaxies projected onto the Coma~I Group (PGC 40692) centered on NGC~4278, 4.8 Mpc from the Virgo core, are blueshifted by 600 \kms\ compared with the mean of a nest of mostly early type galaxies \citep{2011ApJ...743..123K}.  We suspect that the blueshifted galaxies including NGC 4150 belong in the group, although our algorithm does not make that link because of the large velocity differential.  Another instance in our sample of an even more extreme velocity anomaly involves the NGC~1407 Group (PGC 13505) \citep{1993ApJ...403...37G, 1994A&A...283..722Q, 2006MNRAS.369.1375T}, see update in T15$a$.  In that case, it is reasonably certain that NGC~1400 and two dwarfs with extreme blueshifts of 1000~\kms\ are group members. NGC~1407 and NGC~1400 have similar surface brightness fluctuation distance estimates.  With our PGC~40692 and PGC~13505 groups, these galaxies with anomalous velocities are linked by manual override of our algorithm.

Mergers between groups are not uncommon and if the projection geometry is favorable then we can see extreme velocity differentials.  The Centaurus Cluster (PGC~43296) at the extremity of our catalog is a case in point.  Evidently, there is a substantial influx of new members with a line-of-sight velocity differential of 1,500~\kms\ compared with the main structure \citep{1986MNRAS.221..453L, 1997A&A...327..952S}.

\subsection{The Leo Region}

\begin{figure}[!]
\begin{center}
\includegraphics[width=8cm]{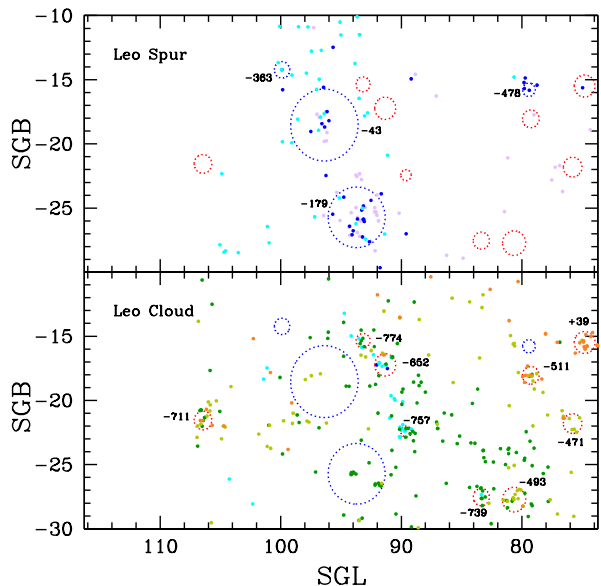}
\caption{
{\it Top.} The foreground Leo Spur. {\it Bottom.} The background Leo Cloud.  Individual galaxies are represented by colors coded as in Fig.~\ref{fig:vsx.map}. Circles illustrate $R_{2t}$ dimensions around the major groups: blue for those in the Leo Spur and red for those in the Leo Cloud.  Peculiar velocities are recorded adjacent the major groups.}
\label{fig:leo.map}
\end{center}
\end{figure}

The availability of precision distances helps address confusion in regions where there are galaxies in projection at substantially different distances but with almost overlapping velocities.  One of these regions is at the convergence on the sky of the structures called the Leo Spur and the Leo Cloud in the Nearby Galaxies Atlas \citep{1987nga..book.....T}.  The foreground Leo Spur and background Leo Cloud are distinct over most of their lengths.  However, in the vicinity of $\ell =235^{\circ}, b = 60^{\circ}$ ($SGL \sim 95^{\circ}, SGB \sim -20^{\circ}$) they overlap.  The well known Leo Group (NGC~3379 = PGC~32256) and NGC~3623/27/28 Group (PGC~34695)  lie near here at 10 Mpc as part of the Leo Spur.   Groups around NGC~3607 (PGC~34426) and NGC~3457 (PGC~32787) lie at $\sim22$~Mpc within the Leo Cloud.  For the most part, galaxies in the Leo Spur have velocities less than 1000~\kms\ in the Local Sheet frame while galaxies in the Leo Cloud have velocities above this limit, but there are important exceptions.  Figure~\ref{fig:leo.map} shows a projection of the region split between the foreground Leo Spur in the top panel and the background Leo Cloud in the lower panel.  The split is informed by distances, if available, with a cut at 16~Mpc.  All galaxies associated with a group are kept together.  In the absence of a distance estimate for a group or individual then a cut is made at a velocity of $V_{LS} = 1000$~\kms.

The circles in the figure represent the radii $R_{2t}$ around the major groups: blue if taken to be in the Leo Spur and red if taken to be in the Leo Cloud.  The circles are repeated in both panels to provide spatial references.  The peculiar velocities, $V_{pec}$, of these major groups are given by the adjacent numbers, where $V_{pec} = V_{LS} - {\rm H}_0 d$, $d$ are weighted group distances, and H$_0 = 75$~\kmsMpc\ is taken as representative of the full data compilation \citep{2016AJ....152...50T}.

The main components of the Leo Spur, with $V_{LS} \sim 650$~\kms\ and distance $\sim 10$~Mpc, have peculiar velocities $\sim -100$~\kms.  The Leo Spur has been studied in some detail \citep{2015ApJ...805..144K}.  As it extends toward us in a layer below the Local Sheet, peculiar velocities rise to $\sim -250$~\kms.  However, our attention here is on the Leo Cloud which manifests very large negative peculiar velocities.  In the region around $SGL \sim 90^{\circ}$, $SGB \sim -20^{\circ}$ observed velocities are $V_{LS} \sim 1000$~\kms, distances are $\sim 23$~Mpc, and peculiar velocities are $\sim -700$~\kms!  Aside from within the virial domain of clusters and the immediate infall vicinity of the Virgo Cluster, these anomalous motions are the largest within the volume of this study.

Numerical action orbit models (Shaya et al. in preparation) provide insight into the confusing situation. Our Local Sheet is streaming at roughly 320~\kms\ toward the Leo structures due to repulsion from the Local Void toward negative SGZ and attraction toward the Virgo Cluster at positive SGY  \citep{2008ApJ...676..184T, 2017ApJ...835...78R}.  We are chasing the Leo Spur toward Virgo in SGY and merging in the SGZ direction, resulting in moderate negative peculiar velocities \citep{2015ApJ...805..144K}. The Leo Cloud is also surging toward the Virgo Cluster but from the back side and, hence, toward us.  In this case the peculiar motions of the Local Sheet and the Leo Cloud are additive and reasonably aligned, creating the 700~\kms\ differential.  The details are a story for another time.  The   
takeaway for the present is that projection confusion can be severe because of velocity streaming but distance information is becoming increasingly useful as a discriminant.

\subsection{Canes Venatici and Ursa Major}

\begin{figure}[!]
\begin{center}
\includegraphics[width=8cm]{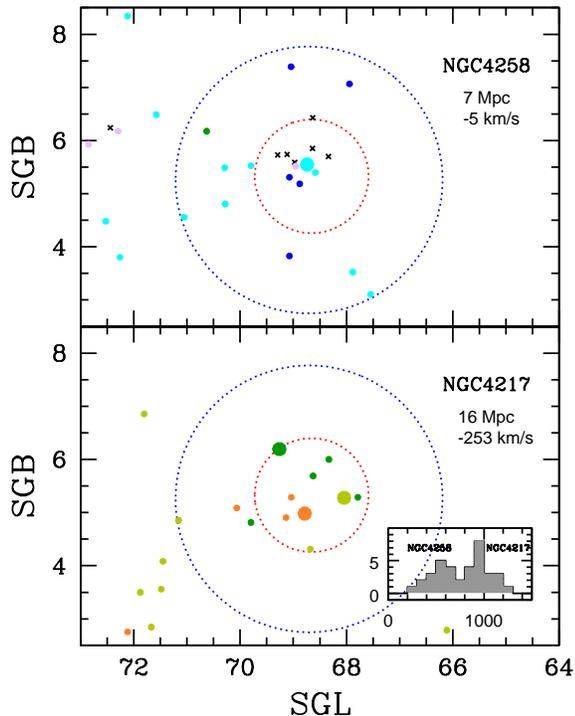}
\caption{
{\it Top.} The foreground NGC~4258 Group. {\it Bottom.} The background NGC 4217 Group.  Velocities $V_{LS}$ in intervals $\pm100$~\kms\ are coded purple (200~\kms), blue (400~\kms), cyan (600~\kms), green (800~\kms), pea green (1000~\kms), orange (1200~\kms).  Velocities are unavailable for the dwarf galaxies represented by black crosses.  The $R_{2t}$ radii for the NGC~4258 and NGC~4217 groups are shown as blue and red dotted circles respectively.  The group peculiar velocities are $-5$~\kms\ and $-253$~\kms\ respectively.  All galaxies in both panels are represented in the velocity histogram inset.}
\label{fig:uma.map}
\end{center}
\end{figure}

Another difficult region lies in the Canes Venatici and Ursa Major constellations. Our Local Sheet projects onto a background band of galaxies extending from the Virgo Cluster along the supergalactic equatorial plane.  Figure~\ref{fig:uma.map} illustrates the problem where the confusion is most acute.   The NGC~4258 Group (PGC~39600) to the foreground at 7 Mpc projects directly onto the NGC~4217 Group (PGC~39241) at 16 Mpc. This latter group is part of what has been called the Ursa Major Cluster \citep{1996AJ....112.2471T} but which, in fact, is a filament \citep{2013MNRAS.429.2264K}.  There are large relative peculiar velocity differences, with the foreground essentially at rest in the Hubble flow while the background Ursa Major filament has a peculiar velocity toward us of $\sim -270$~\kms. As a consequence the observed radial motions almost overlap in spite of substantial differences in distances. 
Nevertheless, a cut at $V_{LS} \sim 700$~\kms\ reasonably discriminates between the foreground and background galaxies.   All individual distance measures for galaxies with $V_{LS}<700$~\kms\ are less than 13 Mpc while all distance measures for galaxies with $V_{LS}>800$~\kms\ are greater than this value.  There are two galaxies in the interval $700-800$~\kms\ with distance estimates: one placed to the foreground at 9 Mpc and the other to the background nominally at 23 Mpc.

\section{Special Case: Nearest 10 Mpc}
\label{sec:10Mpc}

The nearest 10  Mpc deserves special attention.  The inventory of dwarf galaxies is good within this distance \citep{2013AJ....145..101K} and then falls off rapidly in completeness.  Ten Mpc is the limit for a reliable tip of the red giant branch distance determination with a single orbit observation with Advanced Camera for Surveys with Hubble Space Telescope.  Distances accurate to $5-7\%$ are available for about 400 galaxies within this limit,\footnote{http://edd.ifa.hawaii.edu} roughly 60\% of the known galaxies with $M_B<-12$.  Completion is $85-90\%$ within 4 Mpc.

\begin{figure*}
\begin{center}
\includegraphics[width=16cm]{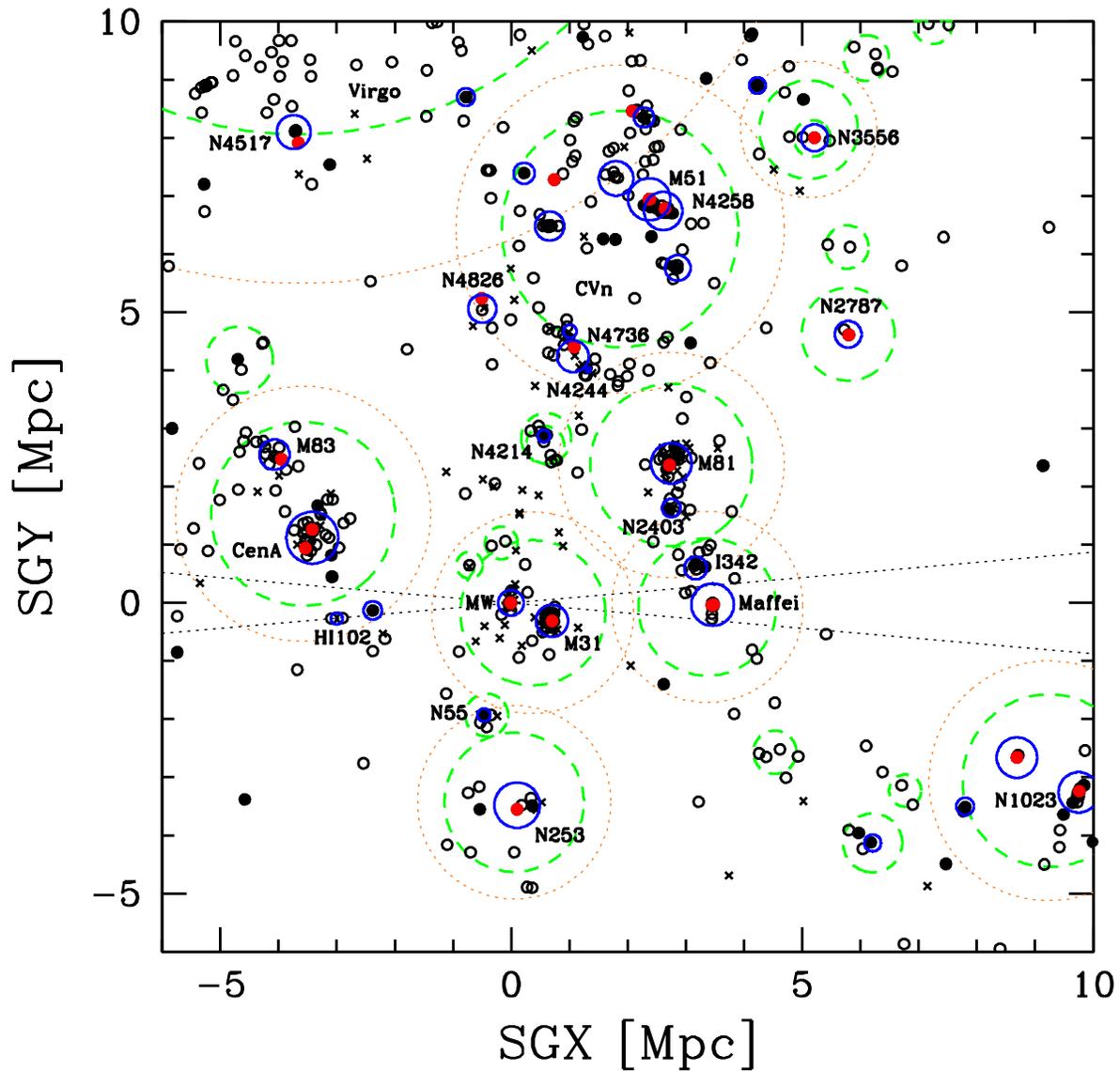}
\caption{
Distribution of galaxies in the Local Sheet.  The equatorial slice in supergalactic coordinates is $3>{\rm SGZ}>-2$ Mpc.  Symbol colors and shapes distinguish galaxy masses: red $>10^{12}$~\Msun, filled black $10^{11}-10^{12}$~\Msun, open $10^{10}-10^{11}$~\Msun, cross $<10^{10}$~\Msun.  Blue circles delineate $R_{2t}$ radii of larger groups, dashed green circles represent the $R_{1t}$ domains, and dotted orange circles approximate the zero energy surfaces.  Galactic latitudes $\pm 5^{\circ}$ are coincident with the dotted wedge. }
\label{fig:6mpc}
\end{center}
\end{figure*}

These accurate distances close to home permit the clear delineation of groups.  It is convenient for display purposes that most nearby galaxies lie within a thin plane we call the Local Sheet \citep{2008ApJ...676..184T}.  Galaxies in a slab extending $-6$ to $+10$~Mpc from our position along the supergalactic cardinal directions in the plane and $-2$ to $+3$~Mpc in projection are represented in Figure~\ref{fig:6mpc}.  Individual galaxies with inferred masses greater than $10^{12}$~\Msun\ are highlighted in red.  Solid blue circles indicate the $R_{2t}$ domains of all groups with at least two members while green dashed circles identify $R_{1t}$ domains. 

The availability of accurate distances has led to surprises, resolving the confused projection of entities that qualify as dwarf associations, including NGC~55 and NGC~300 at 2.0 Mpc onto the NGC~253 Group at 3.6 Mpc, and NGC~4214 with companions at 2.9 Mpc projected onto the NGC~4736 Group at 4.5 Mpc.  In both cases, velocities alone are not a discriminant but distance measures clearly identify the collections of dwarfs to be distinctly to the foreground \citep{2006AJ....132..729T}.  In another case, distances make clear that there are distinct Cen~A and M83 groups (at 3.7 and 4.7 Mpc respectively) although there is projection overlap and the velocity regimes are indistinguishable (T15$a$).

It was just mentioned that accurate distances reveal the existence of two collections of dwarfs in projection against confusing larger structures.  Our terminology needs to be clarified.  These two dwarf collections and several others nearby \citep{1981AJ.....86..340G} were called dwarf associations by \citet{2006AJ....132..729T}.  However now we would prefer to reserve the name ``association" for the more precise concept of the volume enclosing the infall region around a collapsed halo.  The collections of dwarfs need to be evaluated according to this restrictive criterion. The properties of these entities are intriguing.  Their existence suggests the frequent presence of halos in the range $10^{11} - 10^{12}$~\Msun.  Indeed, it is logical that halos exist in this range and these dwarf assemblies must be their manifestation. \citet{2012AstBu..67..135M} have drawn attention to such entities.  However, if so, the mass-to-light ratios are large, an anticipated condition from comparisons of the galaxy luminosity function and theoretically anticipated halo mass function \citep{2002ApJ...569..101M, 2003MNRAS.345..923V}.  

\begin{figure}[!]
\begin{center}
\includegraphics[width=8cm]{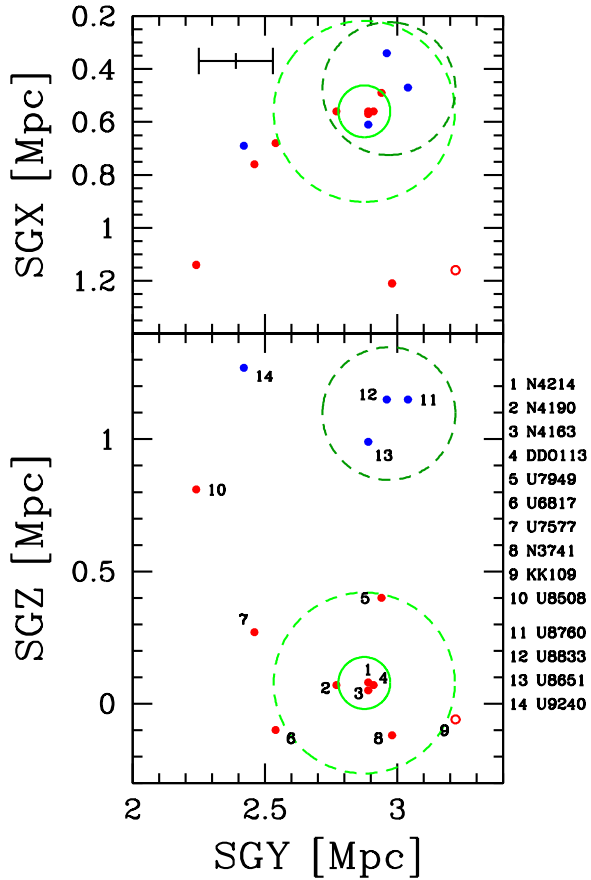}
\caption{
Two projections in supergalactic coordinates. Galaxies nearest the dwarf associations 14+7 and 14+8 are colored red and blue respectively.  Solid and dashed green circles indicate projected $R_{2t}$ and $R_{1t}$ surfaces respectively.  Distance errors reflect predominantly along the SGY axis and the error bar in the upper left corner indicate a typical value.  The open symbol is located on the basis of a velocity because the object lacks a distance measure.}
\label{fig:2dw}
\end{center}
\end{figure}

\noindent
{\it Around NGC 4214.}
In Section~\ref{sec:general_group} we specified the dependence between mass and light below $10^{10}$~\Lsun\ as $M_v^{exp}/L_{K_s} = 32 L_{10}^{-0.5}$~\Msun/\Lsun.  This formulation is poorly constrained because there are few plausibly collapsed halos in this low luminosity regime with a multiplicity of members necessary for a virial estimate of mass.  Our best case at present is the entity called 14+7 in \citet{2006AJ....132..729T}.  The volume that includes this entity along with the adjacent 14+8 is shown in Figure~\ref{fig:2dw}, an enlargement of the region in Fig.~\ref{fig:6mpc} around $SGX\sim0.5$, $SGY\sim3$~Mpc.  The dominant galaxy in 14+7 is NGC~4214 = PGC~39225 ($M_B=-17.2$, roughly midway in luminosity between LMC and SMC).  There are four galaxies within a core region that can be considered a common halo (PGC~39225 plus companions PGCs 38881, 39023, and 39145; respectively NGCs 4214, 4163, 4190, and DDO 113).  The velocity dispersion of these four galaxies is 49~\kms\ giving an estimated virial mass of $1.2\times10^{11}$~\Msun.  The mass to $K_s$ band light ratio is 84~\Msun/\Lsun.  Our formulation based on luminosity predicts a similar mass of $1.2\times10^{11}$~\Msun\ and halo domain $r_{2t}=130$~kpc.  The infall domain characterized by $r_{1t}=460$~kpc captures three more galaxies (PGCs 37050, 40904, 43129; respectively UGCs 6817, 7577, 7949).

The nearby 14+8 consists of three dwarfs (PGCs 48332, 49158, 49452; respectively UGCs 8651, 8760, 8823) that are less than 200~kpc distant from each other and a fourth at 600~kpc.  However our formulation predicts individual $r_{2t}$ halo values of only $\sim 70$~kpc, hence that each dwarf is the sole resident of its distinct halo.  The three adjacent dwarfs in 14+8 probably lie in a common infall zone of radius $\sim 300$~kpc while the fourth escapes.  The remaining three dwarfs in Fig.~\ref{fig:2dw} must also be loners.   

\begin{figure}[!]
\begin{center}
\includegraphics[width=8cm]{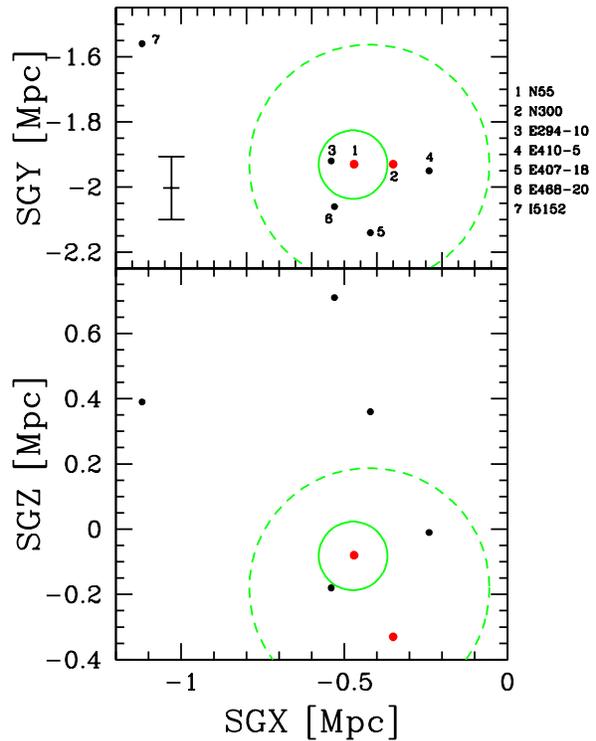}
\caption{
Two projections in supergalactic coordinates of the NGC~55 $-$ NGC~300 region.  The solid circles encloses the NGC~55 halo while the dashed circles enclose the association infall domain.  The error bar in the upper panel is indicative of distance uncertainties, predominantly aligned with the SGY axis.}
\label{fig:n55}
\end{center}
\end{figure}

\noindent
{\it Around NGC 55.}
As seen in the projections of Figure~\ref{fig:n55}, NGC~55 (PGC~1014) and NGC~300 (PGC~3238) lie 277 kpc from each other in the company of two other dwarfs, ESO 410-05 (PGC~1038)  and ESO 294--10 (PGC~1641).  Only PGCs 1014 and 1641, separated by 120 kpc, are plausibly in a common halo but the foursome are probably within a common infall zone attributed a mass of $1.6\times10^{11}$~\Msun\ and an extent $r_{1t}=500$ kpc by our algorithm.  A crude application of the virial theorem to the four galaxies with dispersion 44~\kms\ gives mass $4\times10^{11}$~\Msun.   A fifth galaxy ESO 407-18 (PGC~71431) lies near the edge of the infall region and two others lie beyond (ESO 468-20 and IC 5152; respectively PGC~69468 and PGC~67908).

\begin{figure}[!]
\begin{center}
\includegraphics[width=8cm]{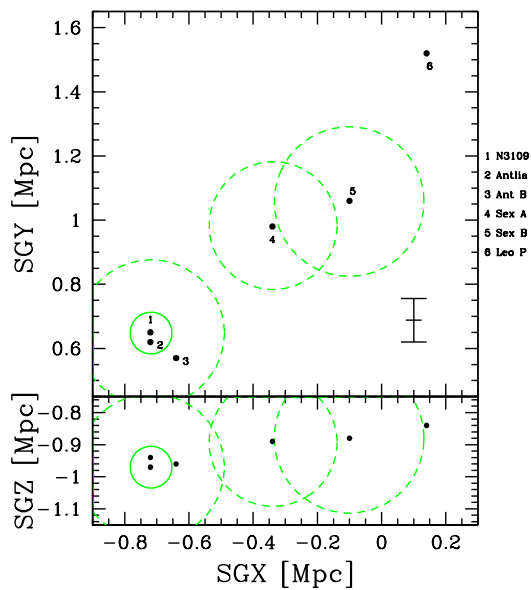}
\caption{
Two projections in supergalactic coordinates of the NGC~3109 region.  NGC~3109 and the two Antlia dwarfs are probably caught in a common halo. Sex~A and B are likely falling together.}
\label{fig:n3109}
\end{center}
\end{figure}

\noindent
{\it Around NGC 3109.}
The nearest collection of dwarfs at 1.4~Mpc, 14+12 \citep{2006AJ....132..729T}, includes NGC~3109 (PGC~29128), Antlia (PGC~29194), Antlia~B (PGC~5098252), Sex~A (PGC~29653) and B (PGC~28913), and possibly Leo~P (PGC~5065058) \citep{2015ApJ...812..158M} in a roughly linear alignment \citep{2013A&A...559L..11B}.  See the distribution in Figure~\ref{fig:n3109}.  NGC~3109 and Antlia are undoubtedly in a common halo, and possibly also Antlia~B \citep{2015ApJ...812L..13S}, but only these three.  Infall zones anticipated by our algorithm are illustrated in the figure.  The two Sextans galaxies must be in separate halos but are probably in a common infall domain.  It would take more mass that ascribed by our algorithm for these galaxies to merge with NGC~3109.  Leo~P is alone.

A complement to the identification of collections of dwarf galaxies has been the clarification that dwarf galaxies rarely lie in extreme isolation, although there definitely are instances.  The lack of dwarf galaxies in under dense regions \citep{2001ApJ...557..495P, 2011AJ....141..204K, 2017ApJ...835...78R} is an extension of the well known problem of under-representation of dwarfs in groups \citep{1999ApJ...522...82K, 1999ApJ...524L..19M}.  

In summary of this discussion of dwarfs, the volume within 10 Mpc provides a currently unique opportunity to study small galaxy groups and the faint end of the halo mass function because only nearby do we have the membership discrimination provided by accurate distance information.  We have to contend, though, with the fact that while the coverage is extensive it is not complete.  In any event, our results are consistent with the claim by \citet{2017NatAs...1E..25S} based on the statistics of dwarf correlations at larger distances that bound multiple systems of dwarfs are rare.

Returning to Fig.~\ref{fig:6mpc}, one sees an inventory that should be complete of the major halos in the nearby portion of the Local Sheet (identified by the blue circles) and of the approximate domains of infall of associations.  The group around NGC 4258 (PGC~39600) hosts the largest nearby association (although the single dominant galaxy that bequeaths the association name is M51 = PGC~47404).  This association lies just outside the capture zone of the Virgo Association, seen encroaching in the upper left of the figure.

It is seen that the $R_{1t}$ radii around the M81 and Maffei groups tentatively overlap.  It would not be surprising if these entities are destined to merge.  
In the standard $\Lambda$CDM model, the surface for ultimate collapse lies $\sim 40\%$ beyond the current turnaround surface  \citep{2008A&A...488..845P}.  These `zero energy' domains are identified by the dotted orange circles around the major constituents in Fig.~\ref{fig:6mpc}.
Our Local Association is ultimately likely to merge with the Maffei and M81 associations.   
The CVn Association is vulnerable to ultimate collapse into the Virgo Cluster.  Will our Local-Maffei-M81 assembly be pulled along?  A more sophisticated analysis accounting for tides is required.
The Centaurus A and NGC 253 associations may drift off alone.

\section{Group catalog}
\label{sec:group.catalog}

We constitute an all-sky sample of 15,004 galaxies in groups with $V_{LS}<3500$ km s$^{-1}$.  There are 1,536 groups with two or more members involving 7,714 galaxies. The radial velocities of the galaxies inside these groups might not all be less than 3500 km s$^{-1}$ interval due to the internal velocity dispersion of groups. In addition, we catalog 7,290 singles, groups of one known galaxy. Table \ref{tab:group_stat} gives statistics on the number of galaxies in our all-sky sample and the identified groups. The columns in this table are (1) the area of the sky, separating northern and southern galactic hemispheres. (2) The number of galaxies in each area. (3) The number of galaxies with measured distances in the CF3 catalog. (4) Number of identified groups (at least two members) with central radial velocity less than 3500 km s$^{-1}$. (5) Number of grouped galaxies. (6) Number of groups with at least 5 members with the total number of galaxies in (7). Columns (8) and (9) give the number of groups that include at least one galaxy with {\it Cosmicflows-3} distance information, and the 
total number of galaxies in these groups.

\begin{table*}
\fontsize{8}{12}\selectfont
\caption{\small The number of galaxies in our sample in northern and southern Galactic hemispheres. Columns (2), (4), (6) and (8) report the number of galaxies or groups satisfying specified criteria, and (3), (5), (7) and (9) give the corresponding total number of galaxies. }
\label{tab:group_stat}
\medskip
\begin{tabular}{c |c c | c c | c c | c c   }
\hline
\hline
Sample & \# of all  & \# of galaxies  & \# of   & \# of   & \# of groups & \# of  & \# of groups   & \# of  \\
       & galaxies   & w. distances & groups  & galaxies        &  $N\geqslant5$ & galaxies & w. distances & galaxies\\ 
(1) & (2) & (3) & (4) & (5) & (6) & (7) & (8) & (9) \\
\hline
North & 10489 & 2025 & 1066 & 5842 & 234 & 3850 &  538 & 4406 \\
South & 4515  & 986  & 470  & 1872 &  75 & 934  &  260 & 1327 \\
Total & 15004 & 3011 & 1536 & 7714 & 309 & 4784 &  798 & 5733 \\

\hline
\end{tabular}
\centering
\end{table*}

In Table \ref{tab:group_catalog}, we present the 10 most luminous groups, as an abbreviation of the full table available on-line. Column (1) gives PGC1, the group ID in our catalog based on the PGC ID of the most luminous galaxy of the group from which the clustering process is started. 
Column (2) gives PGC1+, the ID of the group parent association. Column (3) gives the number of galaxies in each group. Columns (4-7) give the coordinates of group centers in the galactic and supergalactic coordinate systems. In column (8), $K_s$ is the total $K_s$-band apparent magnitude of the group based on the integration of the $K_s$-band luminosity of all galaxies in group. From the 2MRS11.75 catalog, we have at least the coverage of all galaxies brighter than $K_s=11.75$ mag. Column (9) shows the logarithm of the $K_s$-band absolute luminosity of groups in solar units. Columns (10) and (11) provide the radial velocity of groups calculated by unweighted averaging the radial velocity of their member galaxies. $V_h$ is the heliocentric radial velocity of group and $V_{LS}$ is its radial velocity relative to the Local Sheet. Column (12) gives the number of group members with measured distances in the CF3 catalog. Column (13) gives the distance of groups based on the weighted averaged distance moduli of any members available in the {\it Cosmicflows-3} distance catalog. Column (14) provides the percentage error in the distance estimation in column (13).  In recognition of systematics, a minimum to errors is set at 3\%. Columns (15) and (16) contain alternative velocity dispersions of the groups. The first, $\sigma_L$ is the dispersion based on the mass estimate from total luminosity, Equation \ref{eq:M_v.Sigma}. The second, $\sigma_V$ is the line-of-sight velocity dispersion of the group calculated based on the radial velocities of its members, Equation \ref{eq:sigma-p}. Column (17) gives the second turnaround radius of groups based on their luminosity mass, Equation \ref{eq:R_Mv}. Column (18) gives the projected virial radius of groups based on Equation \ref{eq:virial_unweight}. Column (19) presents the logarithm of the mass of groups in solar units, derived from the $K_s$-band luminosity by using $M/L$ ratios given in Equation \ref{eq:M_L}. Alternatively, column (20) gives the dynamical mass of groups, calculated from virial radius, R$_g$, and dynamical velocity dispersion, $\sigma_V$, using Equation \ref{eq:virial_mass}.

Table \ref{tab:galaxy_catalog} provides the catalog of individual galaxies with their group ID. The PGC ID of galaxies is presented in column (1). Column (2) gives the common name of galaxies. Columns (3-8) give galaxy coordinates in Equatorial, Galactic and Super Galactic reference frames. Column (9) gives the morphological classification of each galaxy in the Third Reference Catalogue format \citep{1991trcb.book.....D}, with negative numbers assigned to early type galaxies and positive numbers to late type galaxies. Columns (10) and (11) give the magnitudes of the galaxies in $B$- and $K_s$-band. Log($K_s$), V$_h$, V$_{LS}$, D and eD in columns (12-16) have the same meaning as in Table \ref{tab:group_catalog}. Column (17) gives PGC1, the parent group ID of each galaxy. For the isolated galaxies, PGC and PGC1 IDs are the same.

Table \ref{tab:association} provides the catalog of group associations. Column (1) gives the ID of each association. PGC1+ is the ID of the most luminous galaxy group inside the association. SGL, SGB, log($K_s$), V$_{LS}$, D and eD in columns (2-7) have the same meaning as in Table \ref{tab:group_catalog}. Column (8) gives the estimated 3 dimensional first turnaround radius of each association. Column (9) is the logarithm of the integrated mass of all member groups, where mass is derived from the $K_s$-band luminosity (column (18) of Table \ref{tab:group_catalog}).\footnote{The table KT Groups in the Extragalactic Distance Database (http://edd.ifa.hawaii.edu) combines the information in Tables 2$-$4; each row gives information on an individual galaxy along with information regarding its group and association.  While the on-line tables are fixed at publication, the table at the Extragalactic Distance Database is expected to receive updates.}

\begin{figure}[!]
\begin{center}
\includegraphics[width=3.2in]{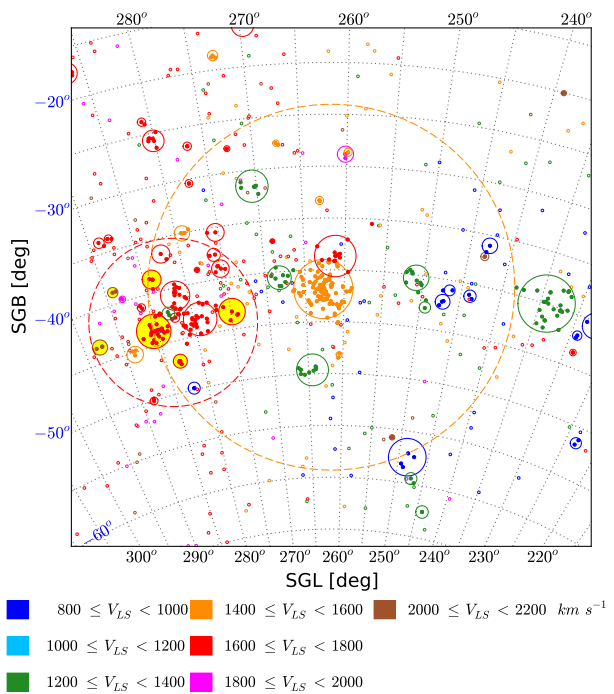}
\caption{Projection of the distribution of galaxy groups in the vicinity of the Fornax Cluster.  Galaxies within groups bounded by solid circles are given a color coded by the group mean velocity.  The Fornax Cluster with dominant galaxy NGC~1399 (PGC~13418), near the center in orange, is the focus of an association bounded by the largest dashed circle.  The NGC~1407 group and association (PGC~13505) overlaps in projection but apparently lies sufficiently to the background to be distinct.  The major groups in this latter association are shaded yellow. 
}
\label{fig:fornax.map}
\end{center}
\end{figure}

The spatial distribution of groups around the Virgo Cluster was displayed in Figures \ref{fig:vir68} and \ref{fig:vsx.map} and discussed in Section~\ref{sec:virgo}.  The region around the Fornax Cluster is seen in Figure \ref{fig:fornax.map}.   The considerable number of galaxies at higher supergalactic longitude than the main cluster belong to what has been called the Eridanus cloud \citep{1987nga..book.....T}.  It has been argued that the entire Eridanus cloud is destined to fall into the Fornax Cluster \citep{2006MNRAS.369.1351B}.  With distances presently available, we find the NGC~1407 (PGC~13505) group and association to be slightly disjoint to the background but with association boundaries only a couple of Mpc apart. Accordingly, it is not expected that the Fornax and NGC~1407 structures will combine in the next Hubble time but are likely within a common zero-energy surface and will ultimately merge.

\begin{figure*}[!]
\begin{center}
\includegraphics[width=7.5in]{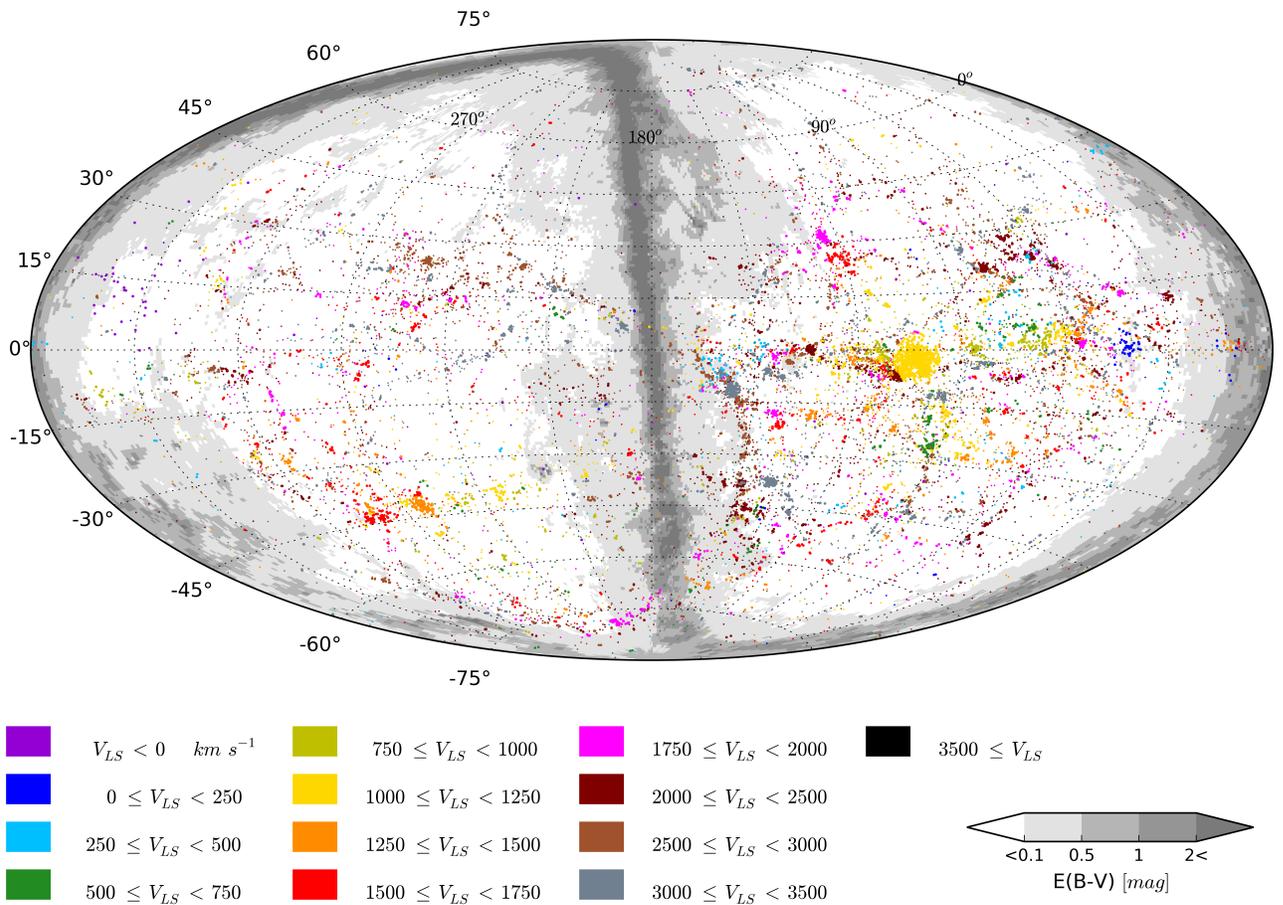}
\caption{
An Aitoff supergalactic projection of all identified groups with $V_{LS}<3500$ km s$^{-1}$. Right and left halves of this diagram correspond to the Galactic northern and southern part of the sky, respectively. Galaxies in groups are represented with the same color corresponding to the group radial velocity.  Contours of Galactic extinction are drawn from the dust maps of \citet{1998ApJ...500..525S}.
 }
\label{fig:AllSky.map}
\end{center}
\end{figure*}

The spatial distribution of all groups and individual galaxies with $V_{LS}<3,500$~\kms\ is presented in Figure \ref{fig:AllSky.map}.   Major features live near the equator in supergalactic coordinates and most are to the right in the north Galactic hemisphere.  The Virgo Cluster is right-center; Fornax Cluster is lower-left; the Local Void is upper-center; the Centaurus Cluster is near center, to the galactic north of the Milky Way plane.  Prominent filaments emanate from Centaurus, one toward 10 o'clock that emerges south of the galactic plane to link up with the Telescopium-Grus Cloud in the Nearby Galaxies Atlas \citep{1987nga..book.....T}, a pair fork toward the Virgo Cluster at 3 o'clock, and a major filament heads toward 6 o'clock to meet the Antlia Cluster then arcs behind the galactic plane near the south supergalactic pole.  This latter feature has been called the Hydra Wall  \citep{1998lssu.conf.....F} north of the galactic plane, the Puppis filament \citep{1992A&A...266..150K} as it passes through the plane, and the Antlia-Hydra and Lepus clouds that bracket the plane \citep{1987nga..book.....T}.  Pomar\`ede et al. (2017) refer to the feature as the Centaurus-Puppis-PP filament because they note that it can be followed all the way from the Centaurus Cluster to the Perseus-Pisces supercluster at 5,000~\kms.  See the latter reference for further descriptions of the major features of nearby large scale structures.

We do not provide alternative group identifications from the literature in our tables.  However, a convenient mechanism for making such linkages is provided within the Extragalactic Distance Database.  As an example, consider the most closely comparable catalog by \citet{2011MNRAS.412.2498M}.  Group linkages can be made by simultaneously opening our `KT Groups' catalog and the `MK Groups' catalog.  Only pertinent columns in each catalog need be selected.  Similarly (and simultaneously) other catalogs with group information can be opened.  The `Tully 3000' catalog contains the groups identified by \citet{1987ApJ...321..280T, 1988ngc..book.....T}.  Cross references with the `2MRS1175 North/South Groups' catalogs provide links with 2MASS based group studies \citep{2007ApJ...655..790C, 2011MNRAS.416.2840L, 2015AJ....149..171T}.

\section{Group Properties}
\label{subsec:investigations}

The individual properties of group members can be summed or averaged to give group parameters. The line-of-sight velocity dispersions and virial radii, $R_g$ of groups can be calculated using Equations \ref{eq:sigma-p} and \ref{eq:virial_unweight}. However, we are confronted with small groups with low numbers of members, and therefore we need to use a method to minimize the effect of uncertainties. According to \citet{1990AJ....100...32B}, bi-weight location and scale estimators are more efficient in working with poor statistics and are less affected by outliers. They are also more robust when working with diverse population characteristics. Accordingly, we use bi-weight statistics to determine the line-of-sight velocity dispersion and central radial velocity of groups. According to the \citet{1990AJ....100...32B} bi-weight method, the gravitational radius of each group is calculated based on the angular distribution of its galaxies and an estimation about the group distance. In the roughly one half of groups without a measured distance we  use the Hubble 
distance to convert angular size to physical size. The virial mass of groups, $M_v$, is then calculated using Equation \ref{eq:virial_mass}.

\begin{figure}[!]
\centering
\hspace{-0.2in}
\subfigure
{
    \label{fig:sub:a}
    \includegraphics[width=3.1in]{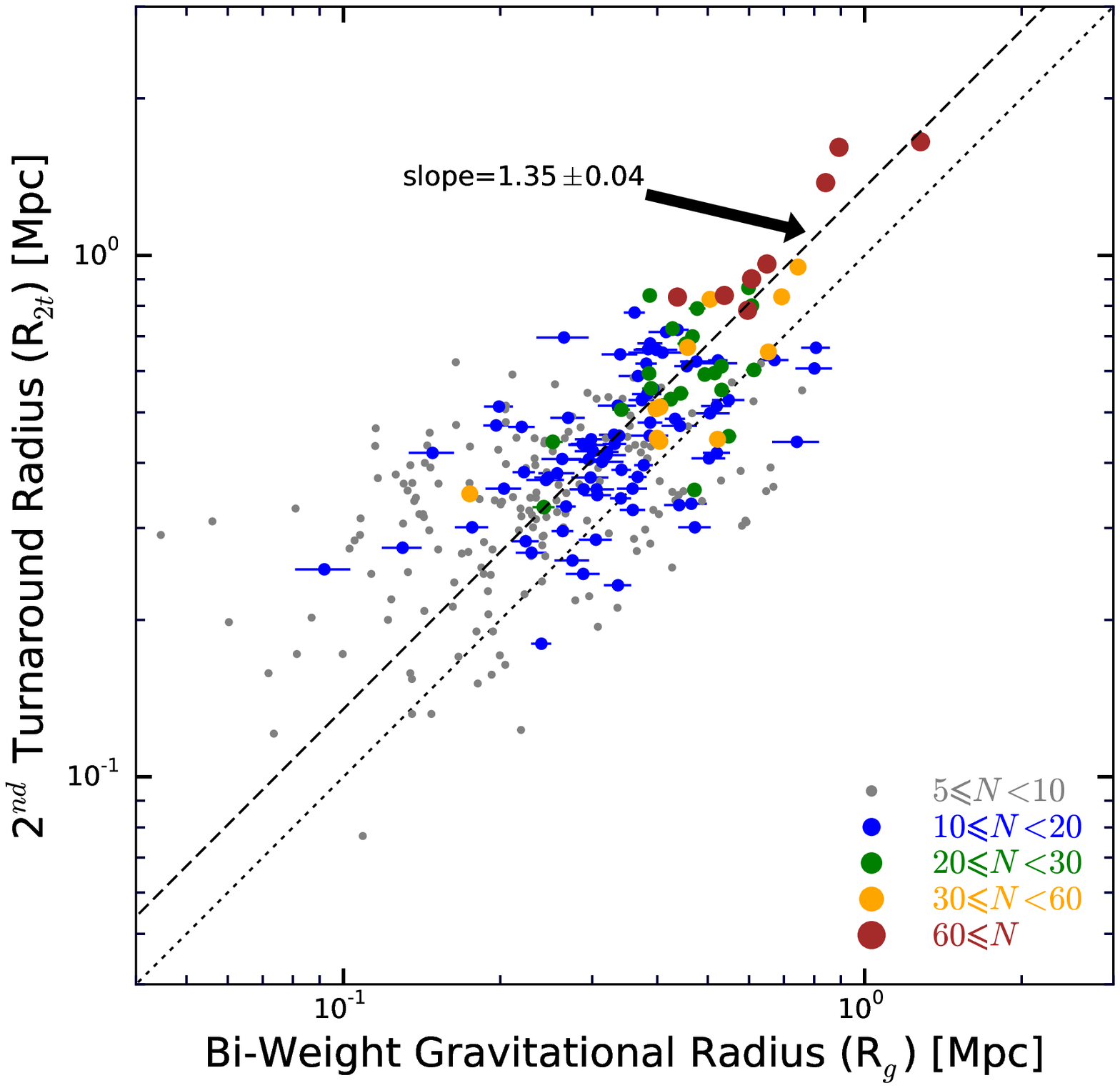} 
}
\hspace{-0.1in}
\subfigure
{
    \label{fig:sub:b}
    \includegraphics[width=3.1in]{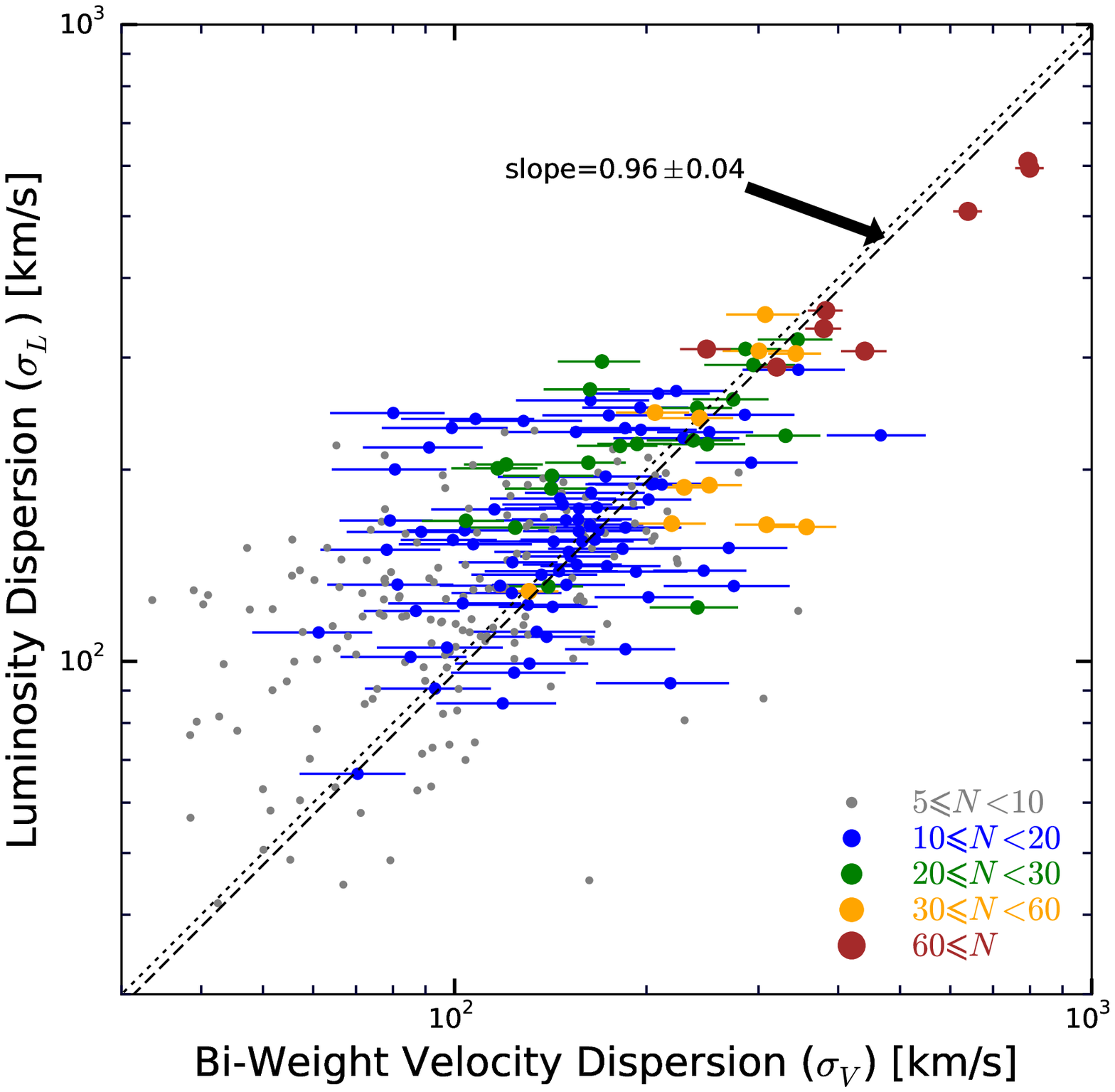} 
}
\caption{{\it Top:} 2$^{nd}$ turnaround radius calculated using Eq. \ref{eq:R_Mv} vs. bi-weight gravitational radius. {\it Bottom:} Velocity dispersion of groups based on their luminosity using Eq. \ref{eq:M_v.Sigma} vs. their bi-weight velocity dispersion. Dotted lines trace equality between parameters. Dashed lines represent the fitted lines for groups with $N \geq 20$, weighted by estimated errors.}
\label{fig:bi-weight2} 
\end{figure}

Figure \ref{fig:bi-weight2} illustrates comparisons between the scales and velocity dispersions of groups. The dotted line shows the equality between two measures and the dashed line is the best fitted line for groups with at least 20 members, i.e. $N \geq 20$. The top panel plots the  second turnaround radius that is derived based on the total $K_s$-band luminosity versus the group bi-weight gravitational radius. The second turnaround radius, $R_{2t}$, is 35\%$\pm$4\% larger than the gravitational radius, somewhat larger than the 27\% found with the more distant 2MRS11.75 sample of T15$b$. In the bottom panel, the luminosity dispersion that is determined based on Equation \ref{eq:M_v.Sigma} is plotted versus the bi-weight velocity dispersion of groups. As seen, both measures are statistically the same for groups with large numbers of galaxies, with bi-weight velocity dispersions $4\%\pm4\%$ larger for groups with $N \geq 20$..  However for groups with fewer galaxies the bi-weight dynamical velocity dispersion tends to be smaller than the velocity dispersion estimated from luminosity. This offset is attributed to the poor statistics in measuring the dynamical velocity dispersion from only the line-of-sight component of velocities and for a small number of objects.

\begin{figure}[!]
\begin{center}
\includegraphics[width=3.1in]{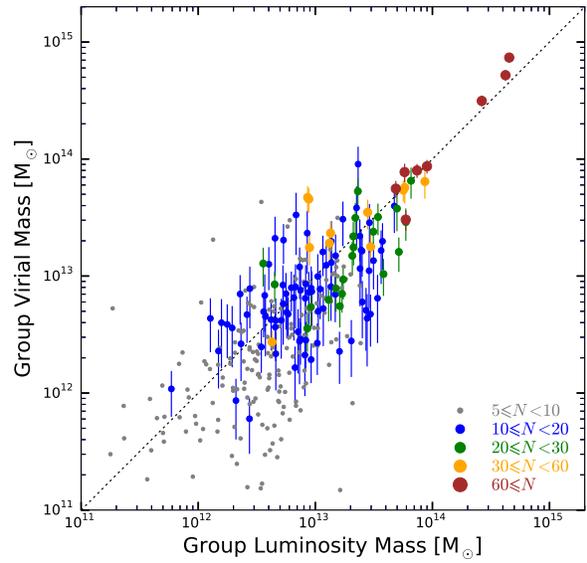} 
\caption{
Estimated virial mass of groups based on Equation \ref{eq:virial_mass} vs. 
luminosity mass. To estimate the virial mass, we use the bi-weight radii and 
velocity dispersion of groups. The dotted line shows equality between the mass 
estimations. Color coding is based on the number of galaxies in each identified 
group.  Errors in virial mass predominate, especially in groups with few members.}
\label{fig:bi_mass}
\end{center}
\end{figure}

Figure \ref{fig:bi_mass} compares the virial mass of groups using Equation \ref{eq:virial_mass}, when all input parameters are determined using bi-weight formulations, and their luminosity mass. For massive groups, the alternate estimations are consistent: for groups with at least 20 members virial masses are $5\%\pm12\%$ larger.  For less massive groups, luminosity masses tend to be larger. The trend can be understood from the statistics of small numbers and particularly because only the radial component of velocities is observed.  Galaxies with close separations in radial velocity will be more favorably captured than member galaxies with extreme radial velocity differences.

\subsection{Group Crossing Times}
\label{subsec:crossingtime}

\begin{figure}[!]
\begin{center}
\includegraphics[width=3.1in]{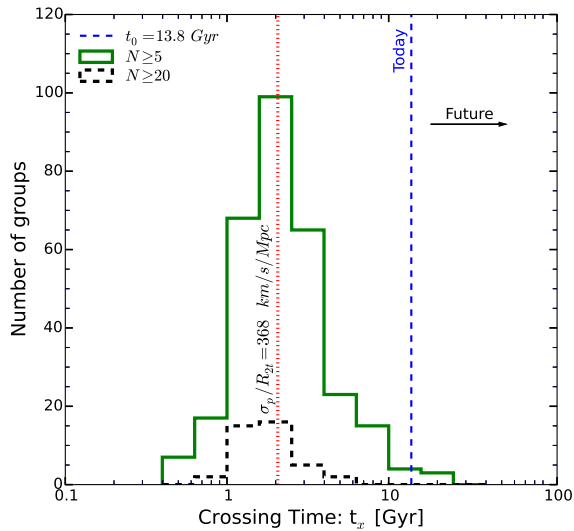}
\caption{
Distribution of crossing time, $t_x$ for the identified groups. $t_0=13.8$ Gyr 
is the age of the universe. The green histogram shows the distribution of 307 groups with at least 
5 galaxies, whereas the black histogram represents those 40 groups 
with at least 20 galaxy members. The vertical red dotted line shows 
the crossing time following from  Eq. \ref{eq:sig_R2t}.} 
\label{fig:crossing.time}
\end{center}
\end{figure}

To evaluate whether it is reasonable that our groups have collapsed, we compare the time they need to form a bound system within the age of universe, $t_0=13.8$ Gyr. The standard crossing time for each group is defined as \citep{1987ApJ...321..280T}

\begin{equation}
t_x \equiv  \frac{r_g}{\sigma_{3D}} \approx \frac{1.05\sqrt{3/2}R_{2t}}{\sqrt{\alpha}\sigma_p} , 
\label{eq:crossingtime}
\end{equation}
where $\alpha = 2.5$. Figure \ref{fig:crossing.time} shows the distribution of crossing time, $t_x$ for the identified groups. Out of 307 groups with at least 5 members, the crossing times of 4 groups (1\%) are larger than the age of universe, which we can attribute to the low number of galaxies in these groups and the happenstance of low velocities in the radial projection. For comparison, the black dotted distribution is plotted for groups with at least 20 galaxies. The vertical red dotted line is at the 2~Gyr group crossing time determined based on Equation \ref{eq:sig_R2t} and is seen to be consistent with the peak of both plotted distributions.

\section{The Group Mass Function}
\label{sec:massFunction}

The construction of a halo mass function lead to a considerable surprise.  The halo mass function was built from all groups, including groups of one galaxy, within 40 Mpc, excluding the zone within $15^{\circ}$ of the Galactic plane.  Masses are derived from group luminosities, a more stable parameter than virial masses especially in groups with few members.
Commonly the mass function is constructed using the familiar $V/V_{max}$ procedure \citep{1968ApJ...151..393S}, but this method can give distorted results if clumping is extreme and velocities are a poor measure of distance as is the case with a very local sample.  Instead, we determine the halo mass function in discrete shells in distance then normalize the shell counts over mass domains of reliable overlap.  Outer shells provide better statistics at higher mass but are more incomplete at lower mass.  Inner shells have the inverse advantage and disadvantage.

\begin{figure}[!]
\begin{center}
\includegraphics[width=8cm]{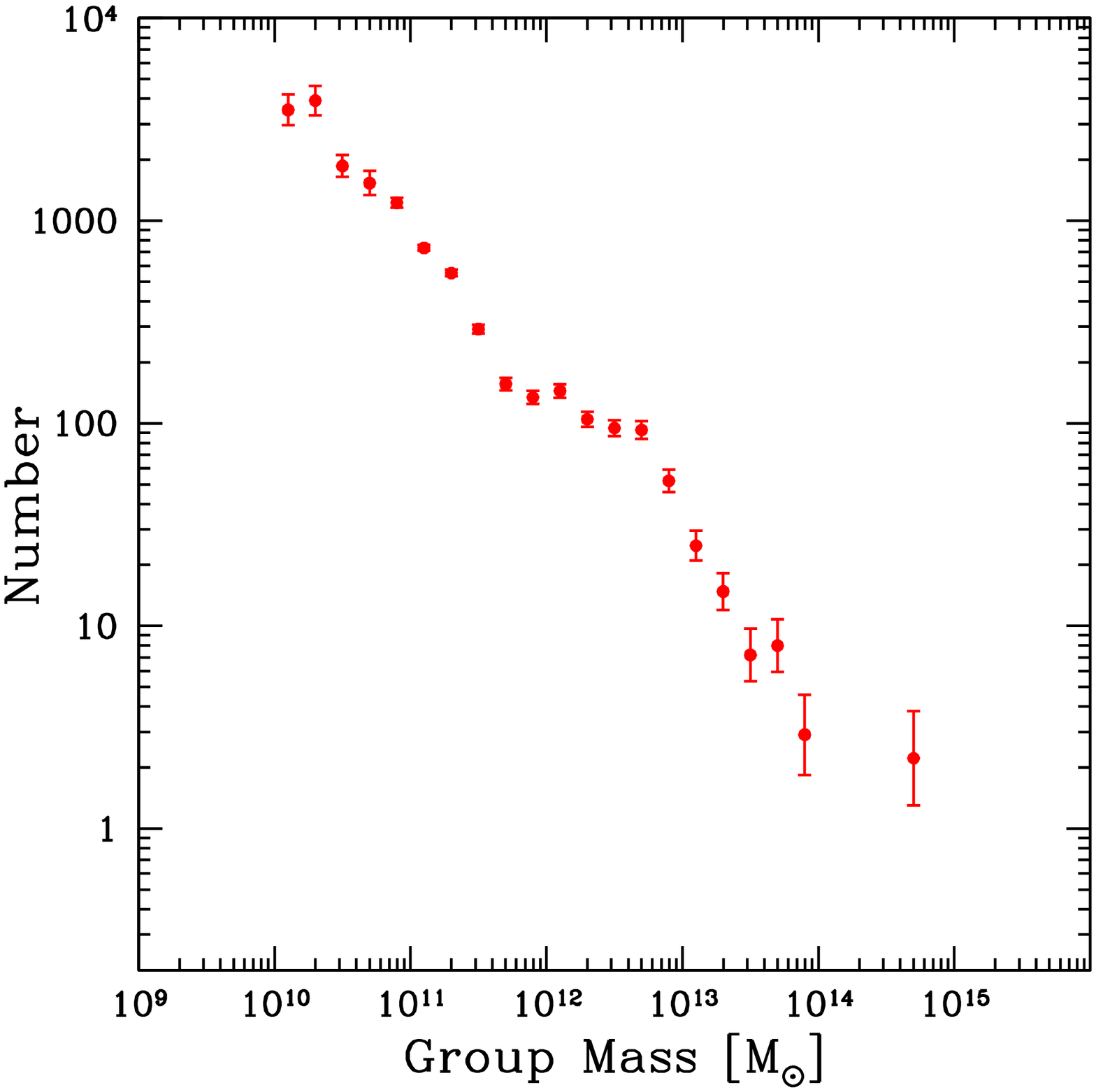}
\includegraphics[width=8cm]{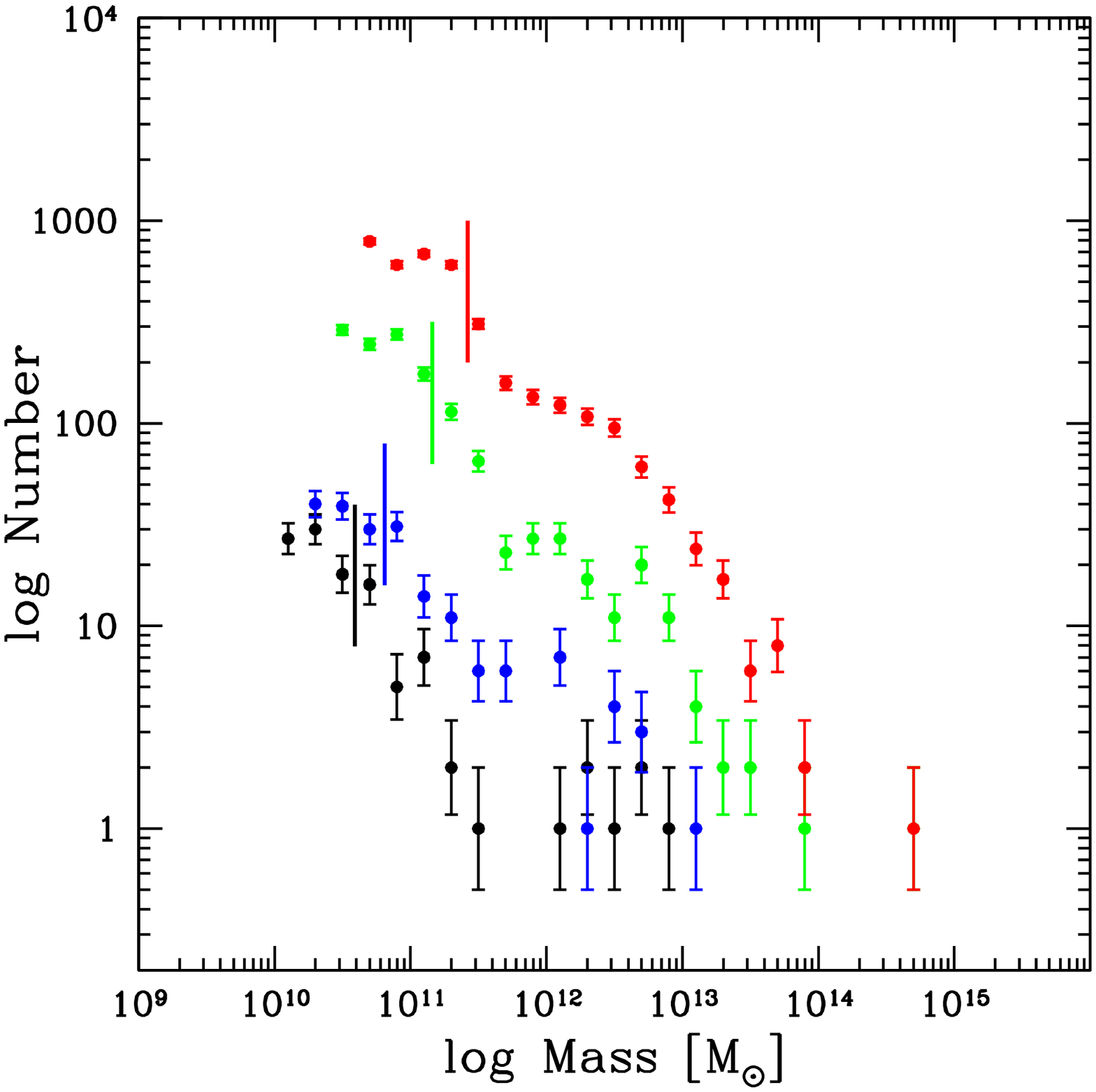}
\caption{The group halo mass function for the sample of galaxies within 3,500~\kms.  In the bottom panel the mass function is separated into 4 bins: within 6 Mpc (black), $6-10$ Mpc (blue), $10-20$ Mpc (green), and $20-40$ Mpc (red).  The separate mass functions are truncated at low mass at the onset of incompletion. The vertical bars associated with each bin indicate the luminosity mass of a single galaxy at the faint limit of the 2MRS11.75 catalog and the distance extreme of the bin.  The bins are normalized and weighted averaged to give the mass function in the top panel.
}
\label{fig:massfn}
\end{center}
\end{figure}

In all, 6176 groups are available, separated into four distance intervals: 143 within 6 Mpc, 239 between 6 and 10 Mpc, 1594 between 10 and 20 Mpc, and 4200 between 20 and 40 Mpc.  The mass functions in these separate bins is seen in the bottom panel of Figure~\ref{fig:massfn}, in each case cut at the low mass end by incompletion.  The vertical lines indicate 2MRS11.75 completeness limits.  Our inventory should include all galaxies brighter than $K_s = 11.75$, modulo the loss of low surface brightness flux.  Associated masses from Eq.~\ref{eq:M_L} at the distance extremes of the four bins lie at the masses of the vertical lines. 

The merged mass function is given in the top panel.  The count normalization is with respect to the counts in the outer (20-40 Mpc) distance bin.  The statistical uncertainties per mass bin represented by the $1\sigma$ error bars are modest between $10^{11}$ and $10^{13}$~\Msun, with uncertainties growing to $10^{14}$~\Msun\ and incompleteness evident below $10^{10}$~\Msun.  An obvious feature of the mass function seen in Fig.~\ref{fig:massfn} is a jog at $\sim10^{12}$~\Msun.

\begin{figure}[!]
\begin{center}
\includegraphics[width=8cm]{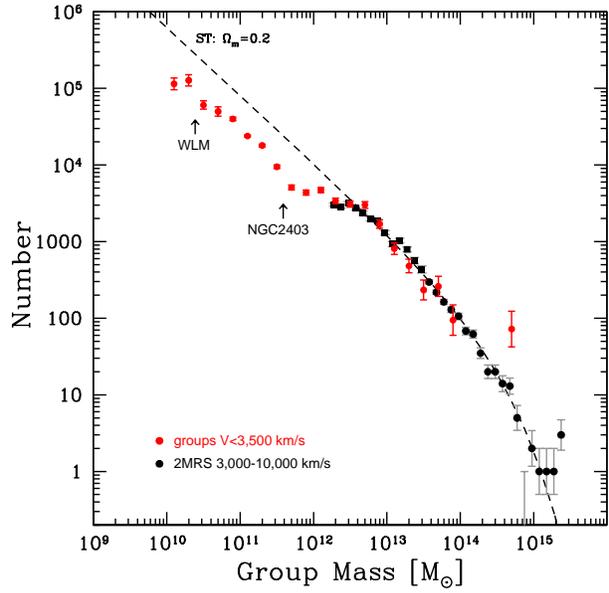}
\caption{The group halo mass function extended.  The mass function derived from the present sample of galaxies within 3,500~\kms, in red, is combined with the mass function determined from the 2MASS11.75 sample of galaxies with velocities between 3,000 and 10,000~\kms, in black.  The dashed curve is a representation of the theoretical Sheth-Torman halo mass function. The locations in mass of the NGC 2403 and WLM halos are identified as examples near the jog from the Sheth-Torman function and the faint limit of the sample, respectively.}
\label{fig:massfn2}
\end{center}
\end{figure}

Figure~\ref{fig:massfn2} merges the mass function of the current analysis with that found with the 2MASS11.75 redshift survey sample of \citet{2012ApJS..199...26H} in the velocity interval 3,000 to 10,000~\kms\ (T15$b$).  This latter sample provides good coverage of the mass function between $10^{12}$ and $10^{15}$~\Msun.  Here the count normalization is with respect to the more distant 2MASS11.75 construction.

The evidence for a jog at $\sim10^{12}$~\Msun\ is reinforced in Fig.~\ref{fig:massfn2}.  Already there was a hint of this feature in Fig. 14 of T15$b$, although the apparent bending of the mass function there was too close to the low mass limit to be convincing.  In this new analysis, the mass range around $10^{12}$~\Msun\ is easily accessible in all our distance bins and the apparent inflection in this mass range is seen in all bins.

\citet{1974ApJ...187..425P} developed a formalism describing the expectation halo mass function with the hierarchical growth of bound structures in the evolving universe.  In T15$b$ the preferred fit to the 2MASS11.75 mass function was drawn from the theoretical variation of  \citet{1999MNRAS.308..119S} assuming the cosmological parameter $\Omega_m=0.2$ with a flat cosmic topology.  That fit is shown in Fig.~\ref{fig:massfn2} as a dashed curve.

The superposition of the theoretical curve draws attention to the acuteness of the jog at $10^{12}$~\Msun.  Is the departure from the theoretical expectation due to an observational artifact or the manifestation of a physical process?

We have investigated several observational concerns.

\noindent
(i) Above $10^{12}$~\Msun\ most halos have multiple galaxy components whereas below this mass it is most common for a halo to host only a single galaxy.

\noindent
(ii) The mass $10^{12}$~\Msun\ is roughly the inflection point in the relationship between halo mass and light, with halos becoming progressively more dark toward both higher and lower mass (Figure~\ref{fig:massTolight}).

\noindent
(iii) Galaxies less massive than $10^{12}$~\Msun\ usually do not have reliable 2MASS $K_s$ magnitudes and the assigned $K_s$ fluxes are inferred from $B$ band magnitudes, themselves often not well known.  

\noindent
(iv) The relationship between $K_s$ flux and (mostly dark) mass has significant uncertainties, especially at the low mass end.

\noindent
(v) Since the low mass regime can only be studied locally and the high mass regime requires inspection of a large volume for decent statistics, the construction of the mass function involves the integration of nested distance shells.

Concerns (ii), (iii), and (iv) are distinct yet coupled in the sense that they each could generate systematics of similar natures.  They each could affect the slope of the observed mass function over significant mass ranges.  However, it is difficult to understand how any of these three concerns can create the relatively sharp jog in the mass function $-$ a flattening then a steepening over intervals of only a factor $\sim 2$ in mass.  Regarding (iii), we have independently confirmed the reasonableness of the conversion scheme from $B$ to $K_s$ magnitudes proposed by \citet{2005AstL...31..299K} as far as can be monitored by available data, but the uncertainties are admittedly large.  The situation could be vastly improved in the future with systematic incorporation of deep infrared or optical photometry from, for example, {\it WISE}, the {\it Wide-field Infrared Survey Explorer} \citep{2010AJ....140.1868W} or {\it Pan-STARRS} \citep{2010SPIE.7733E..0EK}.  Given an estimate of $K_s$ band flux with available information, the conversion to mass remains an issue; the concerns (ii) and (iv) emphasize separate aspects.  The mass function jog occurs near the minimum in our assumed fractional relationship between mass and light.  However our formulation has a broad and smooth transition between high mass and low mass regimes.  We have tried alternative  descriptions with only mild consequences on the jog.

Concern (v) is distinct.  While the interlacing of distance regimes is subject to uncertainties, it is noteworthy that the jog is seen in every distance sub-sample.  The relevant mass regime is well sampled in every distance sub-sample in the current $V_{LS}<3,500$~\kms\ study.

It is concern (i) that would seem most likely to cause the abruptness of the jog.  However this concern invites a physical explanation rather than an artifact.  It is reasonably well established that, if one looks hard enough, halos above $\sim10^{12}$~\Msun\ contain multiple galactic systems.  By contrast, as discussed in Section~\ref{sec:10Mpc}, halos below $\sim10^{12}$~\Msun\ usually only contain a single galaxy.  It is to be appreciated, though, that in halos with around $10^{12}$~\Msun\ most of the luminosity, and by inference mass, is generally in a single galaxy.  The Milky Way, M31, M81, and NGC 253 are all local examples.  Hence, although these groups have satellites, they could effectively be thought of as singles (as such galaxies are, if viewed at a distance).

The question before us is whether the jog in the group mass function near $10^{12}$~\Msun\  represents a real deviation from the Press-Schechter (or Sheth-Torman variant) of the expectation halo mass function of hierarchical clustering theory.  The location and mass of the NGC 2403 Group gives clarity to the sort of entity that lies in the transition zone.  The NGC 2403 galaxy lies within the M81 infall region but with an independent halo containing 3 identified companions.  M33 is a similar galaxy but it, and at least one companion \citep{2013MNRAS.436.2096S}, have recently been subsumed into the M31 halo.  These intermediate size galaxies are not hard to identify.  The theory would suggest that some 55\% of halos in the jog mass regime are not being counted. 
\citet{2011MNRAS.415L..40B} have discussed a `too big to fail' problem with an apparent deficit of large satellites in groups.  Here we draw attention to an apparent deficit of groups themselves in a regime that could be called `way too big to fail'.

\section{Summary}
\label{summary}

The 3,500~\kms\ limit ($\sim 47$~Mpc) embraces the historical Local Supercluster \citep{1953AJ.....58...30D}, reaching at its edge the Centaurus (PGC~43296) and Hydra (PGC~31478) clusters.  The major condensation other than these two is the Virgo Cluster (PGC~41220).  The mass within the Virgo Association, the infall domain, is $\sim 7\times 10^{14}$~\Msun, 60\% of this within the cluster.  The other clusters and associations of note are Fornax (PGC~13418) with an infall mass of $2\times10^{14}$~\Msun, 30\% in the main cluster, Antlia (PGC~30308) with an infall mass of $4\times10^{14}$~\Msun, only 15\% already collapsed, Virgo~W (PGC~39659) with an association of $2\times10^{14}$~\Msun, 60\% in the existing cluster, and NGC~5846 (PGC~53932) with an association of $1\times10^{14}$~\Msun, a third in the present cluster. 

Turning to the cumulative halo mass function, we ask if
the abrupt onset of a deficiency of halos below $10^{12}$~\Msun\ compared with expectations is an observational artifact or is telling us something important about galaxy formation.  A reasonable census at masses below $10^{12}$~\Msun\ requires inclusion of dwarf galaxies with very low surface brightness. It is only nearby, within $<10$~Mpc, that surveys have any sense of completeness at such low masses.  Wide field HI surveys with the Arecibo, Parkes, and Jodrell Banks telescopes are giving an increasingly complete inventory of systems containing cold gas.  Observations of resolved stars with Hubble Space Telescope provide accurate distances.  Dwarfs devoid of gas are common as satellites of large galaxies but as companions they are not markers of separate halos.  There are known examples of gas-poor dwarfs in isolation \citep{2012MNRAS.425..709M,2015MNRAS.447L..85K, 2017MNRAS.464.2281M}.  However, according to our present understanding they are sufficiently rare that their numbers do not significantly impact the halo mass function.

The halo mass function that we are able to construct extends from $10^{15}$~\Msun\ all the way down to $10^{10}$~\Msun.  There are considerable uncertainties at the low mass end.  Candidate halos might be missed.  The translation from luminosity to mass is uncertain.   However, an observed jog in the halo mass function cannot easily be explained as an artifact.  Above $10^{12}$~\Msun\ the mass function behaves as expected from gravitational clustering theory but by $10^{11}$~\Msun\ there appears to be a deficit of halos by a factor $\sim3$. 

We certainly see convincing examples of halos in the range $10^{11}-10^{12}$~\Msun.  The principal galaxies in these halos are intermediate sized systems, extending in morphology from small spirals to large irregulars, galaxies with intrinsic luminosities of $M_B\sim -19$ to $-17$.  At issue is why we find a factor three shortfall from expectations of such systems as the principal galaxies in halos  since galaxies in this luminosity range cannot be missed at distances less than 10 Mpc.



\section*{Acknowledgments}
\acknowledgments

A big thanks to the referee for substantial suggestions.
We are indebted to efforts over many years of Igor and Valentina Karachentsev(a), Dmitri and Lidia Makarov(a), H\'el\`ene Courtois, Luca Rizzi, and Ed Shaya.   
Jaime Forero-Romero is looking at simulations to help us understand the implications of the "way too big to fail" jog in the halo mass function.
We in particular appreciate the effort made to maintain HyperLEDA \citep{2014A&A...570A..13M}.
This research has made use of the NASA/IPAC Extragalactic Database\footnote{\url{http://nedwww.ipac.caltech.edu/}} which is operated by the Jet Propulsion Laboratory, California Institute of Technology, under contract with the National Aeronautics and Space Administration.   



\bibliography{paper}
\bibliographystyle{apj}

\begin{landscape}
\begin{table}[h]
\centering 
\fontsize{8}{12}\selectfont
\caption{\small Catalog of galaxy groups.}
\label{tab:group_catalog}
\setlength{\tabcolsep}{0.5em} 
\begin{tabular}{ c c c c c c c c c c c c c c c c c c c c }
\hline
\hline

PGC1 &   PGC1+ & Mem &     Glon &     Glat &      SGL &      SGB &     K$_s$ &  log(K) &   V$_h$ &  V$_{LS}$ &  N$_D$ &   D &  eD & $\sigma_L$ & $\sigma_V$ & $R_{2t}$ & $R_{g}$ & log(M$_L$) & log(M$_v$) \\
    &          &  \# &      deg &      deg &      deg &      deg &       mag & [L$_{\odot}$]  &    km/s &       km/s &    \#    & Mpc &  \% &       km/s &       km/s &            Mpc &           Mpc & [M$_{\odot}$] & [M$_{\odot}$]  \\
(1) & (2) & (3) & (4) & (5) & (6) & (7) & (8) & (9) & (10) & (11) & (12) & (13) & (14) & (15)  & (16)  & (17)  & (18)  & (19) & (20) \\
\hline 
   
  43296 &   43296 & 191 & 302.2241 &  21.6465 & 156.3251 & -11.5819 &  4.61 & 12.71 & 3407 & 3142 &    59 & 36.96 &   3 &     595 &     800 &   1.612 &  0.893 &   14.624 &   14.717  \\
  46618 &   43296 &  30 & 307.8738 &  19.2865 & 159.6426 &  -6.8008 &  6.07 & 12.11 & 3340 & 3086 &     5 & 37.57 &   7 &     350 &     307 &   0.950 &  0.745 &   13.936 &   13.808  \\
  45174 &   43296 &  34 & 306.0424 &  32.5707 & 146.1941 &  -6.0422 &  6.42 & 11.96 & 3292 & 3059 &     4 & 44.03 &   8 &     307 &     300 &   0.833 &  0.693 &   13.764 &   13.756  \\
  40498 &   43296 &  24 & 297.5648 &  23.0823 & 153.9019 & -15.4648 &  6.48 & 11.92 & 3258 & 2986 &     4 & 28.45 &   6 &     296 &     170 &   0.801 &  0.607 &   13.713 &   13.207  \\
  43557 &   43296 &  22 & 302.9333 &  36.4075 & 141.8602 &  -7.7062 &  6.82 & 11.81 & 3296 & 3066 &     7 & 39.89 &   8 &     267 &     163 &   0.724 &  0.428 &   13.582 &   13.017  \\
  46409 &   43296 &  12 & 307.9495 &  14.9305 & 163.9909 &  -7.4020 &  6.96 & 11.70 & 3148 & 2889 &     4 & 33.18 &   8 &     243 &     174 &   0.659 &  0.399 &   13.459 &   13.045  \\
  45466 &   43296 &  22 & 307.8030 &  38.9261 & 140.2186 &  -3.4022 &  7.42 & 11.48 & 2995 & 2783 &     2 & 38.06 &  12 &     201 &     118 &   0.544 &  0.444 &   13.210 &   12.747  \\
  49424 &   43296 &   7 & 313.8107 &  13.0725 & 166.6082 &  -2.0176 &  7.43 & 11.41 & 2833 & 2591 &     1 & 33.88 &  18 &     189 &     130 &   0.513 &  0.391 &   13.134 &   12.782  \\
  49106 &   43296 &   8 & 312.8342 &  12.8475 & 166.7144 &  -2.9897 &  7.65 & 11.40 & 3066 & 2820 &     1 & 30.06 &  18 &     187 &      97 &   0.507 &  0.204 &   13.117 &   12.243  \\
  41960 &   43296 &   7 & 299.5259 &  22.9197 & 154.5023 & -13.7479 &  7.80 & 11.38 & 3226 & 2957 &     1 & 39.63 &  23 &     184 &     147 &   0.498 &  0.467 &   13.093 &   12.966  \\
      
\hline
\end{tabular}
\begin{flushleft}
 The complete version of this catalog is availabe online.
\end{flushleft}
\end{table}
\end{landscape}

\begin{landscape}
\begin{table}[h]
\fontsize{8}{12}\selectfont
\caption{\small Catalog of galaxies and their groups.}
\label{tab:galaxy_catalog}
\setlength{\tabcolsep}{0.5em} 
\begin{tabular}{ c c c c c c c c c c c c c c c c c }

\hline
\hline
 PGC &      Name &     R.A. &     Dec. &     Glon &     Glat &      SGL &      SGB &   T &     B &    K$_s$ &  log(K) &   V$_h$ &  V$_{ls}$ &      D &  eD  &  PGC1 \\
     &           &   J2000  &   J2000  &     deg  &     deg  &     deg  &     deg  &      &   mag &     mag  & [L$_{\odot}$] &    km/s &     km/s  &    Mpc &  \%  &       \\
 (1) &       (2) &      (3) &      (4) &      (5) &      (6) &      (7) &      (8) &  (9) &  (10) &     (11) &  (12)   &    (13) &      (14) &   (15) & (16) &  (17) \\
\hline

  13418 &  NGC1399  &  54.6212 & -35.4506 & 236.7163 & -53.6356 & 262.5460 & -42.0776 & -4.6 & 10.35 &  6.44 & 11.25 & 1425 & 1283 &  22.08 &  11 &   13418  \\
  13179 &  NGC1365  &  53.4016 & -36.1405 & 237.9564 & -54.5979 & 261.8963 & -40.9741 &  3.2 &  9.83 &  6.59 & 11.19 & 1664 & 1523 &  16.98 &   8 &   13418  \\
  13433 &  NGC1404  &  54.7163 & -35.5942 & 236.9551 & -53.5547 & 262.3357 & -42.1253 & -4.8 & 10.81 &  6.90 & 11.07 & 1873 & 1730 &  18.79 &  10 &   13418  \\
  13318 &  NGC1380  &  54.1149 & -34.9761 & 235.9261 & -54.0585 & 263.2770 & -41.7608 & -2.3 & 10.79 &  6.96 & 11.05 & 1855 & 1716 &  18.62 &   7 &   13418  \\
  13344 &  NGC1387  &  54.2377 & -35.5066 & 236.8246 & -53.9462 & 262.5532 & -41.7605 & -2.9 & 11.70 &  7.52 & 10.82 & 1265 & 1124 &  19.14 &  13 &   13418  \\
  13059 &  NGC1350  &  52.7837 & -33.6284 & 233.6122 & -55.1679 & 265.2873 & -40.8965 &  1.9 & 10.75 &  7.53 & 10.82 & 1918 & 1788 &  19.05 &  14 &   13418  \\
  13333 &  NGC1386  &  54.1925 & -35.9992 & 237.6634 & -53.9660 & 261.9157 & -41.6286 & -0.7 & 12.00 &  8.14 & 10.57 &  907 &  764 &  15.92 &  13 &   13418  \\
  13267 &  NGC1374  &  53.8191 & -35.2263 & 236.3623 & -54.2952 & 263.0067 & -41.4773 & -4.5 & 11.93 &  8.25 & 10.53 & 1475 & 1336 &  20.51 &  10 &   13418  \\
  13609 &  NGC1427  &  55.5808 & -35.3928 & 236.5971 & -52.8547 & 262.4102 & -42.8553 & -4.0 & 11.71 &  8.26 & 10.53 & 1410 & 1265 &  19.95 &  10 &   13418  \\
  13299 &  NGC1379  &  54.0159 & -35.4412 & 236.7220 & -54.1287 & 262.6850 & -41.5954 & -4.8 & 11.77 &  8.32 & 10.50 & 1363 & 1223 &  17.70 &  11 &   13418  \\

\hline
\end{tabular}
\begin{flushleft}
 The complete version of this catalog is availabe online.
\end{flushleft}
\end{table}

\end{landscape}

\begin{table*}[h]
\fontsize{8}{12}\selectfont
\caption{\small Catalog of group associations.}
\label{tab:association}
\setlength{\tabcolsep}{0.5em} 
\begin{tabular}{ c c c c c c c c c }

\hline
\hline

PGC1+ &      SGL &      SGB &  log(K) &  V$_{ls}$ &     D &  eD &   r$_{1t}$ & log(M$_L$)  \\ 
      &     deg  &     deg  &         &    km/s   &   Mpc &  \% &   Mpc    & [M$_{\odot}$] \\
 (1) &       (2) &      (3) &     (4) &      (5)  &   (6) & (7) &  (8) &  (9)  \\
\hline

  43296 & 154.4338 &  -9.7132 & 13.11 & 2975 & 36.82 &   0 & 8.511 &   14.897  \\
  41220 & 104.2597 &  -3.4399 & 12.99 & 1130 & 15.67 &   0 & 7.999 &   14.816  \\
  30308 & 151.4498 & -39.5301 & 12.87 & 2510 & 33.53 &   0 & 6.803 &   14.604  \\
  31478 & 139.6889 & -37.4385 & 12.72 & 3335 & 47.49 &   0 & 6.556 &   14.556  \\
  13418 & 261.6607 & -41.6721 & 12.63 & 1305 & 18.20 &   0 & 5.781 &   14.393  \\
  39659 & 108.3979 &  -7.7374 & 12.39 & 2023 & 31.46 &   0 & 4.968 &   14.196  \\
  53932 & 124.5914 &  28.1223 & 12.39 & 1742 & 26.49 &   0 & 4.745 &   14.134  \\
  57649 & 188.7230 &   5.2738 & 12.32 & 3276 & 41.69 &  14 & 4.459 &   14.053  \\
  46247 & 132.8158 &   0.2109 & 12.29 & 2379 & 28.35 &   3 & 4.332 &   14.017  \\
  37617 &  63.4067 &   2.7217 & 12.28 & 1106 & 17.73 &   3 & 4.217 &   13.981  \\

\hline
\end{tabular}
\begin{flushleft}
 The complete version of this catalog is availabe online.
\end{flushleft}
\end{table*}

\label{lastpage}

\end{document}